\documentclass[12pt,preprint]{aastex}
\shorttitle{H$\alpha$ Variations in Cyg X-1}
\shortauthors{Gies et al.}

\begin{document}

\received{2002 June 14}
\accepted{2002 September 25}

\title{Wind Accretion and State Transitions in Cygnus X-1 \altaffilmark{1}}
\altaffiltext{1}{Based on data obtained at the David Dunlap Observatory, University of Toronto.}

\author{D. R. Gies\altaffilmark{2,3}}

\affil{Center for High Angular Resolution Astronomy,
Department of Physics and Astronomy \\
Georgia State University, Atlanta, GA  30303\\
Electronic mail: gies@chara.gsu.edu}

\altaffiltext{2}{Visiting Astronomer, Kitt Peak National Observatory,
National Optical Astronomy Observatories, operated by the Association
of Universities for Research in Astronomy, Inc., under contract with
the National Science Foundation.}

\altaffiltext{3}{Visiting Astronomer, University of Texas McDonald Observatory.}

\author{C. T. Bolton, J. R. Thomson}
\affil{David Dunlap Observatory, University of Toronto\\
P.O.\ Box 360, Richmond Hill, Ontario, L4C 4Y6, Canada\\
bolton@astro.utoronto.ca, jthomson@crux.astro.utoronto.ca}

\author{W. Huang\altaffilmark{2}, 
        M. V. McSwain\altaffilmark{2,3},
        R. L. Riddle\altaffilmark{2,3,4}, \\
        Z. Wang\altaffilmark{2,5},
        P. J. Wiita,
        D. W. Wingert\altaffilmark{2}}

\affil{Center for High Angular Resolution Astronomy,
Department of Physics and Astronomy \\
Georgia State University, Atlanta, GA  30303\\
Electronic mail: huang@chara.gsu.edu, 
mcswain@chara.gsu.edu, riddle@iastate.edu, wangzx@space.mit.edu,
wiita@chara.gsu.edu, wingert@chara.gsu.edu}

\altaffiltext{4}{Current address: Department of Physics and Astronomy,
Iowa State University, Ames, IA 50011}

\altaffiltext{5}{Current address: Center for Space Research,
Massachusetts Institute of Technology,
70 Vassar Street, Building 37, Cambridge, MA 02139}

\author{B. Cs\'{a}k\altaffilmark{6}, L. L. Kiss\altaffilmark{6}}
\affil{Department of Experimental Physics and Astronomical Observatory, 
University of Szeged, D\'{o}m t\'{e}r 9., Szeged, H-6720, Hungary \\
csakb@physx.u-szeged.hu, l.kiss@physx.u-szeged.hu}

\altaffiltext{6}{Visiting Astronomer, University of Toronto David Dunlap Observatory.}

\slugcomment{ApJ, in press}
\paperid{56241}

%%%%%%%%%%%%%%%%%%%%%%%%%%%%%%%%%%%%%%%%%%%%%%%%%%%%%%%%%%%%%%%

\begin{abstract}
We present the results of a spectroscopic monitoring program 
(from 1998 to 2002) of the H$\alpha$ emission strength in 
HDE 226868, the optical counterpart of the black hole binary, 
Cyg X-1.   The feature provides an important probe of the 
mass loss rate in the base of the stellar wind of the supergiant 
star.   We derive an updated ephemeris for the orbit based 
upon radial velocities measured from \ion{He}{1} $\lambda 6678$. 
We list net equivalent widths for the entire H$\alpha$ 
emission/absorption complex, and we find that there are 
large variations in emission strength over both long  
(years) and short (hours to days) time spans. 
There are coherent orbital phase related 
variations in the profiles when the spectra are grouped by 
H$\alpha$ equivalent width.  The profiles consist of 
(1) a P~Cygni component associated with the wind of the supergiant,
(2) emission components that attain high velocity at the conjunctions
and that probably form in enhanced outflows both towards and 
away from the black hole, and (3) an emission component that
moves in anti-phase with the supergiant's motion.   We argue 
that the third component forms in accreted gas near the 
black hole, and the radial velocity curve of the emission is 
consistent with a mass ratio of $M_{\rm X}/M_{\rm opt} \approx  
0.36\pm0.05$.   We find that there is a general anti-correlation
between the H$\alpha$ emission strength and X-ray flux 
(from the {\it Rossi X-ray Timing Explorer} All Sky Monitor instrument)
in the sense that when the H$\alpha$ emission is strong 
($W_\lambda < -0.5$ \AA ) the X-ray flux is weaker and 
the spectrum harder.   On the other hand, there is no 
correlation between H$\alpha$ emission strength and X-ray flux 
when H$\alpha$ is weak.   We argue that this relationship 
is not caused by wind X-ray absorption nor by the reduction 
in H$\alpha$ emissivity by X-ray heating.  
Instead, we suggest that the H$\alpha$ variations track 
changes in wind density and strength near the photosphere. 
The density of the wind determines the size of X-ray
ionization zones surrounding the black hole, and these in turn control 
the acceleration of the wind in the direction of the black hole. 
During the low/hard X-ray state, the strong wind is fast and the 
accretion rate is relatively low, while in the high/soft state 
the weaker, highly ionized wind attains only a moderate velocity and 
the accretion rate increases.  We argue that the X-ray transitions 
from the normal low/hard to the rare high/soft state are triggered 
by episodes of decreased mass loss rate in the supergiant donor star. 
\end{abstract}


%%%%%%%%%%%%%%%%%%%%%%%%%%%%%%%%%%%%%%%%%%%%%%%%%%%%%%%%%%%%%%%

\section{Introduction}                              % Section 1

Cygnus~X-1 has been one of the most intensively studied X-ray sources
in the sky since its discovery and identification with the
O9.7 Iab supergiant star, HDE 226868 \citep{bol72,web72}.
This system provided the first evidence for
the existence of stellar mass black holes when it was discovered
to be a 5.6 day binary with a massive, unseen companion.
\citet{gie86a} used the spectroscopic orbit \citep{gie82,las98,bro99a}, 
light curve \citep{kem83,kar01}, photospheric line broadening, 
and a range in the assumed degree of Roche-filling of the supergiant star
to obtain mass estimates of $M_{\rm opt} = 23 - 43 ~M_\odot$
and $M_{\rm X} = 10 - 21 ~M_\odot$.  \citet{her95} derived  
physical parameters for the visible supergiant based upon a 
spectroscopic analysis of the line spectrum, and they adopted
a system inclination of $i=35^\circ$ (based upon published estimates) 
to arrive at mass estimates of 18 and $10 ~M_\odot$ for the supergiant
and black hole, respectively (the derived masses scale as $\sin^{-3} i$
over the probable range of $i= 30^\circ - 40^\circ$; \citet{gie86a,wen99}).  
The X-ray source in Cyg~X-1 is powered mainly by accretion from
the strong stellar wind of the supergiant star \citep{pet78,kap98a}. 
In fact, Cyg~X-1 probably represents a situation 
intermediate between pure, spherical wind accretion and accretion by
Roche lobe overflow.  Observations of the optical emission lines
\citep{gie86b,nin87} indicate that the
wind departs from spherical symmetry and that there exists an enhanced
wind flow (or ``focused wind'') in the direction of the companion.
The intense X-ray emission is believed to be produced close to
the black hole in an accretion disk that emits soft X-ray photons
and in a hot corona that inverse-Compton scatters low energy
photons to higher energies \citep{lia84,tan95}.
Ultraviolet radiation from close to the black hole has been
detected through High Speed Photometer observations with the 
{\it Hubble Space Telescope} \citep{dol01}.   
Radio jets were recently discovered in
Cyg~X-1 \citep{sti01,fen01} indicating a collimated
outflow with a speed in excess of $0.6 c$, so that Cyg~X-1 joins the
group of Galactic {\it microquasars}, small scale versions of active
galactic nuclei \citep{mir99}.  The target is also a candidate $\gamma$-ray 
transient source \citep{gol02}; the $\gamma$-rays are probably created 
through inverse-Compton scattering in the jets.

Cyg~X-1 is generally found in either a low/hard state (the more common
case of low 2-10 keV flux and a hard energy spectrum) or a high/soft
state (in which the soft X-ray flux increases dramatically and the  
spectrum softens; \citet{zha97,zdz02}).  Every few years Cyg~X-1 makes a transition
from the low/hard to the high/soft state, and it remains in this
active state for weeks to months before returning to the low/hard
state.  The last well documented high/soft state occurred in 1996
\citep{bro99b}, and in 2001 September Cyg~X-1 once again
entered a high/soft state in which it still remains 
at the time of writing (2002 September).  
This recent transition into the high/soft state was accompanied by a 
sudden decrease in radio flux \citep{poo01} (the opposite
of the radio brightening that accompanied the return 
to the low/hard state in 1971 and that led directly to
the identification of the visible star associated with the X-ray source; 
\citet{hje71}).  There are many theories about the 
causes of the transitions \citep{cha95,pou97,mey00,wen01,you01,rob02} 
which generally relate to the physical conditions of the gas surrounding 
the black hole.   For example, \citet{esi98} describe the transitions in 
terms of an advection-dominated accretion flow (ADAF) model in which 
the transitions are related to changes in the inner radius of the 
geometrically thin, optically thick, Keplerian disk.   
In the usual low/hard state, the inner disk radius is relatively
large, but during the high/soft state the inner radius extends inwards
close to the last stable orbit around the black hole.  The transition to 
the high/soft state is generally believed to be the result of a moderate 
increase in the mass accretion rate.  

However, there is no clear observational evidence
available to support the claim of enhanced mass transfer during the high/soft state.
In fact, the evidence collected so far hints that the supergiant
mass loss rate may actually decline during the high/soft state.  
\citet{wen99} present a model for the X-ray light curve of Cyg~X-1 
during the prolonged low/hard state based upon the accumulated data 
from the {\it Rossi X-ray Timing Explorer Satellite} (RXTE) 
All Sky Monitor (ASM) instrument (see also \citet{kar01}).
They find that the decreased X-ray flux observed when the supergiant
is in the foreground can be explained by the X-ray absorption caused by
the wind outflow from the supergiant.  However, during the high/soft state
the orbital variation in X-ray flux disappears, and Wen et al.\ argue that
the wind absorption declines because of increased photoionization of
the wind by the stronger X-ray source and a decrease in the wind density
(implying a factor of 2 decrease in the mass loss rate).
\citet{vol97} found that the H$\alpha$ emission associated
with the wind loss also declined during the 1996 high/soft state.
Taken at face value, these observations suggest that the wind mass
loss actually declines during a high/soft state, in contradiction to the
theoretical expectations.  

In this paper we report on multiple year observations of the H$\alpha$ 
emission line in HDE 226868 (\S2) which we find to show significant long 
term and short term variability (\S3), presumably reflecting changes in the 
wind close to the supergiant.   The RXTE/ASM instrument has provided 
continuous X-ray flux measurements of the binary throughout this period, 
and we show that temporal variations in the X-ray flux are broadly  
anti-correlated with the H$\alpha$ emission strength (\S4).   We discuss the 
implications of this result, and we suggest that the state transitions 
may result from changes in the wind velocity that are related to 
the ionization state of the wind (\S5).  

%%%%%%%%%%%%%%%%%%%%%%%%%%%%%%%%%%%%%%%%%%%%%%%%%%%%%%%%%%%%%%%

\section{Observations and Orbit}                    % Section 2

The 115 optical spectra were made from three different sites 
over the period from 1998 August to 2002 May, and we list
in the final column of Table~1 the specific telescope and
spectrograph combination associated with each observation. 
The majority of the spectra were obtained with the Kitt Peak National
Observatory 0.9-m Coude Feed Telescope in conjunction with 
a program on SS~433.  Details about these observations 
are given in \citet{gie02}.   During the observing runs in 1998 and 
2000, we used the long collimator, grating B (in second order with order
sorting filter OG550), camera 5, and a Ford $3072\times 1024$ CCD (F3KB) detector.   
This arrangement produced a resolving power, $R=\lambda / \Delta\lambda = 9530$, 
and covered a range of 829 \AA ~around H$\alpha$.   The 
KPNO Coude Feed runs in 1999, however, relied upon the short collimator, 
grating RC181 (in first order with a GG495 filter to block higher orders), 
and camera 5 with the same F3KB detector, and this set up 
provided lower resolving power, $R = 4630$, but broader spectral 
coverage (1330 \AA ).   
We usually obtained two consecutive exposures of 30 minutes duration
and co-added these spectra to improve the S/N ratio.   
We also observed the rapidly rotating A-type star, $\zeta$~Aql,
which we used for removal of atmospheric water vapor lines.
Each set of observations was accompanied by numerous bias, flat field,
and Th~Ar comparison lamp calibration frames.

\placetable{tab1}      % Table 1 - Journal of Optical Spectroscopy

We also obtained in 1999 several high dispersion echelle 
spectra using the University of Texas McDonald Observatory 2.1-m telescope 
and Sandiford Cassegrain Echelle Spectrograph \citep{mcc93}.
The detector was a Reticon $1200\times 400$ CCD (RA2)
with $27 \mu$m square pixels which recorded 27 echelle orders
covering the region blueward from H$\alpha$ with a resolving power of $R=42000$.
The exposure times were typically 30 minutes. 
The observations from 2001 to 2002 were made 
with the University of Toronto David Dunlap Observatory (DDO)
1.9-m telescope.   The DDO spectra were obtained with the
Cassegrain spectrograph and a Thomson $1024\times 1024$ CCD.
All but one of these spectra were made with a 1800 grooves mm$^{-1}$ 
grating that yielded a resolving power of $R=11200$.  
The exposure times were usually 20 minutes. 

The spectra were extracted and calibrated
using standard routines in IRAF\footnote{IRAF is distributed by the
National Optical Astronomy Observatories, which is operated by
the Association of Universities for Research in Astronomy, Inc.,
under cooperative agreement with the National Science Foundation.}.
All the spectra were rectified to a unit continuum by the fitting
of line-free regions using the IRAF task {\it continuum}
(and in the case of the echelle spectra, the resulting orders were
then linked together using the task {\it scombine}).
The removal of atmospheric lines was done by
creating a library of telluric standard spectra from each run, removing
the broad stellar features from these, and then dividing each target
spectrum by the modified atmospheric spectrum that most closely
matched the target spectrum in a selected region dominated by
atmospheric absorptions.  The spectra from each run were then
transformed to a common heliocentric wavelength grid.  
The higher resolution spectra were smoothed to the nominal,
two pixel resolution of the lower resolution spectra obtained with 
the KPNO Coude Feed and the RC181 grating.   This degradation 
in resolution is inconsequential for the analysis of the broad 
structures in H$\alpha$ that we describe here.  

\placetable{tab1}      % Table 1 - Journal of Optical Spectroscopy

We discuss below the spectral variations related to orbital phase, 
and in order to determine an accurate ephemeris for the orbit, 
we decided to measure radial velocities using the photospheric 
line \ion{He}{1} $\lambda 6678$.  The red wing of this feature is 
blended with \ion{He}{2} $\lambda 6683$, which occasionally is 
filled in with weak, P~Cygni type emission (also observed in 
\ion{He}{1} $\lambda 5876$).   We measured radial velocities 
by fitting a Gaussian function to the central line core in 
order to avoid some of these problems with the line wings, 
but our final results are probably influenced by the varying 
strength of the P~Cygni emission.   We also measured the position 
of the interstellar line at 6613.56 \AA ~in order to correct for 
small differences in the wavelength calibration from the 
different spectrographs (this line is clear of the nearby 
\ion{N}{2} $\lambda 6610$ emission line of the supergiant).  
We measured these differences by cross-correlating 
the profile of the interstellar line in each spectrum with 
the same profile obtained at KPNO on HJD 2,451,425.8079, 
and the measured stellar radial velocities were adjusted 
according to these cross-correlation shifts. 
The final radial velocity measurements are listed in Table~1. 

We made a solution of the orbital elements using the
nonlinear, least-squares fitting program of \citet{mor74}.
All the radial velocities were assigned the same weight in 
computing the solution, and we assumed a circular orbit \citep{gie82}.  
A first trial solution with a fitted period resulted in an orbital period 
that was the same within errors as that derived by 
\citet{bro99a} based upon all the available radial velocity data, 
and since the data used by \citet{bro99a} span a 26 year interval, 
we decided to fix the period at their value, $P=5.599829\pm0.000016$ d. 
We then solved for the remaining parameters:  
the epoch of supergiant inferior conjunction, $T$(IC), 
the semiamplitude, $K$, and systemic velocity, $V_0$.
The measurements and radial velocity curve are shown in 
Figure~1, and our solution is compared to that of 
\citet{bro99a} in Table~2.  This table also lists the r.m.s.\ 
residuals from the fit, $\sigma$, the mass function, $f(m)$, 
and the projected component of the semimajor axis, $a_1 \sin i$.  
The individual observed minus calculated residuals, $(O-C)$, 
and orbital phases, $\phi$, are given in Table~1.   
Our derived epoch is $0\fd043 \pm 0\fd013$ later 
than that predicted by ephemeris of \citet{bro99a}, and we will 
use our new result in what follows.  The semiamplitude is in good
agreement with prior work \citep{gie82,las98,bro99a}, but the 
systemic velocity is somewhat low due to the subtle P~Cygni emission 
in \ion{He}{1} $\lambda 6678$ (which causes the absorption core 
to appear slightly blue-shifted).   Thus, our estimate of 
$V_0$ is probably smaller than the physical systemic velocity 
of the binary.    

\placetable{tab2}      % Table 2 - orbital elements 

\placefigure{fig1}     % Figure 1 - radial velocity curve 

%%%%%%%%%%%%%%%%%%%%%%%%%%%%%%%%%%%%%%%%%%%%%%%%%%%%%%%%%%%%%%%

\section{H$\alpha$ Variability}                     % Section 3
 
The H$\alpha$ feature is regarded as a reliable source for the determination of 
the mass loss rate in luminous hot stars \citep{pul96}.  The line appears as 
an absorption profile in stars with small mass loss rates, and it grows 
into a strong emission profile in stars with well developed winds.  
The emission forms by recombination, and since this process depends on 
the square of the gas density, the H$\alpha$ feature serves as a probe of 
the dense, basal wind near the stellar photosphere.   The observational 
evidence to date demonstrates that the H$\alpha$ emission is also 
time variable, especially in luminous supergiants like the visible star 
in Cyg~X-1.   \citet{kap98b} show an example of large H$\alpha$ emission 
variations over the course of a few days in the O9.5~Ib star, $\zeta$~Ori, 
which presumably result from changes in the wind density and structure. 
\citet{kap97} discuss several cases of cyclical variability in H$\alpha$ 
that is related to the rotational period of the star.   The H$\alpha$ 
feature in Cyg~X-1 appears to be cyclically variable with the orbital 
period \citep{hut74,hut79,nin87,sow98}, but these earlier investigations were 
too limited in time coverage to reveal the large variations that can 
occur that are unrelated to orbital phase.  Both kinds of variability 
offer important clues about the structure and density of the wind outflow 
that is the ultimate source of the accretion fed X-ray luminosity.  

The H$\alpha$ profiles in Cyg~X-1 are often complex in appearance, and 
we decided to begin our analysis by measuring the overall emission strength through a 
simple numerical integration of the line intensity 
(including both emission and absorption portions). 
This equivalent width, $W_\lambda$, was measured over a 40~\AA ~range centered 
on H$\alpha$, and our results are listed in the fifth column of Table~1. 
The measurement errors can be estimated from the values obtained during the 
final KPNO Coude Feed run when Cyg~X-1 was usually observed twice each night. 
Since the emission variations generally occur on longer time scales (days), 
the scatter within a night mainly reflects measurement errors. 
The mean of the differences in equivalent width within each night indicate 
typical measurement errors of $\pm 0.08$ \AA .   

We find that the equivalent width varied from $+0.3$~\AA ~(absorption stronger than emission) to 
$-1.7$~\AA ~(emission much stronger than absorption) over the course of our observations (see Fig.~10 below).   
The variations have both a long term component (significant differences between the 
mean equivalent widths of the individual observing runs) and a rapid component (large 
night-to-night changes).   The emission generally weakened between 1998 and 2002. 
A periodogram search indicated a possible small amplitude variation with a period of 
$156 \pm 8$~d, which is similar to the long X-ray and radio period of $142 \pm 7$~d
found by \citet{bro99b}.  However, our sampling is too fragmentary to 
characterize accurately variations on timescales of months, and we focus here 
on profile variations that are related to orbital phase. 
We found that these were best seen when the profiles were grouped into
samples with similar H$\alpha$ equivalent width.   Figures 2, 3, and 4 illustrate 
the profiles as a function of heliocentric radial velocity and orbital phase for
times between 1998 and 2000 of strong ($W_\lambda < -0.9$ \AA ), 
moderate ($-0.9 < W_\lambda < -0.4$ \AA ), 
and weak ($W_\lambda > -0.4$ \AA ) H$\alpha$ emission, respectively. 
Figure~5 shows the same for the DDO spectra (2001 -- 2002), obtained during
the high/soft X-ray state when the H$\alpha$ emission was generally weak. 
The upper panel in each diagram shows the observed profiles while the 
lower panel is a gray-scale representation of the profiles made through a 
linear interpolation in orbital phase.   The thick line in 
Figure~4 depicts the H$\alpha$ profile observed on 
HJD 2,451,466.6863, just 3 hours prior to the midpoint of 
the recent {\it Chandra} observation of high resolution X-ray spectroscopy \citep{sch02}.   
The emission was weak at that time, and the relatively low column density found 
by \citet{sch02} probably results from the lower density in the wind at that epoch
(or possibly from higher X-ray ionization of the wind; see \S5). 

\placefigure{fig2}     % Figure 2 - Strong H-alpha grayscale 

\placefigure{fig3}     % Figure 3 - Moderate H-alpha grayscale 

\placefigure{fig4}     % Figure 4 - Weak H-alpha grayscale 

\placefigure{fig5}     % Figure 5 - High/soft state H-alpha grayscale 

Each diagram shows the orbital motion of the supergiant star 
as a backwards ``S'' curve in the grayscale image, and there is clear 
evidence of this orbital motion in a blue-shifted absorption and
a red-shifted emission feature (perhaps best seen in the weaker emission 
profiles in Figs.~4 and 5).   This is the characteristic P~Cygni-type shape 
that is the hallmark of mass loss in luminous stars \citep{lam99}, and the association of 
this component with the wind of the supergiant in Cyg~X-1 was demonstrated in earlier 
observations by \citet{nin87} and \citet{sow98}.   The overall shape and 
intensity of the P~Cygni component is comparable to that observed in other 
O-type supergiants \citep{ebb82,kap98b}.  

The next striking feature in these diagrams is the appearance of large 
emission features in the wings of the profiles observed near the orbital conjunction 
phases, $\phi = 0.0$ and 0.5.    These are most evident in the moderate and 
strong emission profiles (Figs.~3 and 2).   Furthermore, we find that the 
central absorption component {\it only} appears at these conjunction phases in the 
strong emission case (Fig.~2).   The light curve solutions for Cyg~X-1 
\citep{gie86a} indicate that the supergiant is close to filling its critical 
Roche surface, and in such a situation the wind is predicted to develop 
enhanced mass outflow in the directions opposite and especially towards the 
black hole companion (the focused stellar wind model of \citet{fri82}). 
We suggest that in the strong emission case the wind is substantially 
enhanced along the axis joining the stars and that at the conjunction phases 
we see the resulting emission with the largest values of projected radial velocity.  
These streams are also seen in partial projection against the disk of the 
star at the conjunctions (which creates the strong absorption seen then) while  
at other phases the emission is seen projected against the sky, resulting in 
an emission filling of the absorption core.   The presence of clumps in the 
stream flowing towards the black hole may partially explain the temporal 
variations we observe and the prevalence of X-ray dips found near phase 
$\phi = 0.0$ \citep{bal00,fen02}.  

All the sets of profiles show some evidence of very blue-shifted absorption 
near and shortly following $\phi = 0.5$, supergiant superior conjunction. 
At this orientation we view the focused outflow in the foreground, and we 
suggest that the blue-shifted absorption forms in the lower density, high latitude 
part of the flow we see projected against the stellar disk while the lower 
velocity blue-shifted emission originates in the denser equatorial part of the 
flow that is mainly projected against the sky.  If we assume that the outflow 
velocity is the same in both the equatorial and our direction at $\phi = 0.5$, 
then the emission component radial velocity is equal to $\sin i$ times 
the absorption component radial velocity (where $i$ is the orbital inclination). 
For $V_r({\rm abs.}) = -380 \pm 50$ km~s$^{-1}$ and 
$V_r({\rm em.}) = -190 \pm 30$ km~s$^{-1}$, we derive an inclination of 
$i= 30^\circ \pm 7^\circ$, which is consistent with the range of values derived 
from the light curve analysis ($i = 28^\circ - 38^\circ$; \citet{gie86a}) and with the 
best fit model for X-ray light curve ($i = 10^\circ - 40^\circ$; \citet{wen99}).   
The assumption of equal outflow velocities 
in these two directions is probably not fully valid.  For example, \citet{fri82} 
find that the wind outflow is faster at high latitudes than in the high density 
equatorial domain, so our estimate of the inclination is probably a lower limit. 

The other important feature in the profile figures is the appearance of a 
blue-shifted emission component near $\phi = 0.25$.  This emission component 
was also found in earlier observations, and it could originate in the 
focused wind flow \citep{nin87,sow98} and/or in high density gas near the black hole
\citep{hut74,hut79}.   In our earlier study of H$\alpha$ \citep{sow98}, we presented 
a simplified scheme to isolate this moving component of emission based upon 
a Doppler tomography algorithm.   The basic assumption is that each profile 
represents the sum of a spectral component that moves with the established 
radial velocity curve of the supergiant (the P~Cygni component discussed above) 
and a second component that moves with a sinusoidal radial velocity curve 
parameterized by a semiamplitude, $K_{\rm em}$, and an orbital phase of 
radial velocity maximum, $\phi_0$.   The tomography algorithm then uses an 
iterative corrections scheme \citep{bag94} to reconstruct the spectral line 
profiles for each component.   We performed tomographic reconstructions over 
a grid of possible $K_{\rm em}$ and $\phi_0$ values to find a solution that 
minimized the differences between the observed and reconstructed composite profiles. 
Since we now have nearly an order of magnitude more spectra than presented 
in \citet{sow98}, we have repeated this procedure with one key difference. 
We showed above that the conjunction phase profiles have shapes that are dominated
by enhanced outflow along the axis joining the stars, and this part of the 
emission will appear broader at conjunctions than at other phases.   
Rather than introducing a third component with phase variable width, we decided 
to include only the non-conjunction phase observations ($\phi =0.1 - 0.4$ 
and $0.6 - 0.9$) in which the emission contributions from the axial outflow will be 
narrow and more concentrated towards the line center (reducing any 
confusion with the moving component seen blue-shifted near $\phi = 0.25$).   

The results of our tomographic grid search for the parameters of the moving 
component are illustrated in Figure~6 in a plot of the root-mean-square 
(rms) residuals of the reconstruction fits.   
There are three regions in the parameter space that offer 
plausible solutions (and all make comparably good fits).  We show the reconstructed 
profiles for each set of parameters in Figure~7.   The first, and in our opinion most
plausible, solution (marked ``A'') occurs for 
$\phi_0=0.79\pm0.04$ and $K_{\rm em}= 218\pm 30$ km~s$^{-1}$ ({\it solid line}). 
The reconstructed profiles show a well developed P~Cygni profile for the component
moving with the supergiant, while the second component is a single emission peak 
that moves with a large but approximately anti-phase motion (probably associated with 
gas near the black hole; see below).  This radial velocity curve is shown 
as an ``S'' curve in the grayscale images of Figures 2 -- 5. 

\placefigure{fig6}     % Figure 6 - RMS image for range of solutions

\placefigure{fig7}     % Figure 7 - trial reconstructions 

A second solution (marked ``B'') is found near 
$\phi_0=0.81\pm 0.06$ and $K_{\rm em}=56 \pm 30$ km~s$^{-1}$ ({\it dotted line}), and
in this case the anti-phase moving component ({\it right hand panel}) is broader 
and double-peaked.   This is close to the solution advocated in \citet{sow98} 
($\phi_0=0.86$ and $K_{\rm em}=68$ km~s$^{-1}$) for which 
the Doppler shifts resemble those observed in the 
\ion{He}{2} $\lambda 4686$ emission line \citep{gie86b,nin87}.
Nevertheless, the adoption of this solution for the moving component in H$\alpha$ 
is problematical.  First, the double-peaked structure of the reconstructed moving component
suggests that it is a numerical artifact of a forced solution in which the slower 
moving emission is compensating for deeper blue absorption in the supergiant component. 
Second, the enhanced axial outflow discussed above is not treated in this simple 
two-component approach, and this emission will tend to favor low velocity amplitude 
solutions (indeed, if the conjunction phase profiles are included in the 
reconstructions, this solution becomes more favorable).   For both of these 
reasons, we believe this second, low velocity solution is untenable.  

The final solution (marked ``C'') is another numerical artifact in which the moving component
moves in phase with the supergiant but with a higher velocity 
($\phi_0=0.23 \pm 0.05$ and $K_{\rm em}=185 \pm 50$ km~s$^{-1}$, {\it dashed line}). 
In this case, the moving component is a complex absorption line that 
modulates a double-peaked emission profile for the supergiant.  
We can rule out this last solution since there is almost certainly no physical 
scenario consistent with such properties.  

The simple, two-component model for the H$\alpha$ profiles suggests 
that the feature results from wind loss from the supergiant 
(the primary's P~Cygni profile) and gas emission with an 
anti-phase radial velocity curve.   We can use the derived kinematical 
parameters of the second component to explore its location in the system. 
The orbital geometry is sketched in Figure~8 in which we plot the 
system dimensions assuming that the supergiant is close to Roche filling
\citep{gie86a}.   The projected radial velocity vector derived for 
the moving component is the sum of an orbital velocity vector (dependent 
on location in the system) and a possible gas flow vector.  The two 
simplest possibilities for these vector combinations are shown in the 
right side of the diagram.  In the first case, we assume that the 
emitting gas cloud is centered along the axis joining the stars and that 
a small vector component of gas flow from the supergiant to the 
black hole is responsible for the slight phase shift of the emission. 
This might correspond, for example, to high density gas from the 
focused wind encountering the disk around the black hole.  Then, 
the black hole will be situated close to or somewhat beyond this emission 
location, implying a mass ratio of $q=M_{\rm X}/M_{\rm opt} \leq 
K_{\rm opt}/(K_{\rm em} \sin (-\phi_0))$.  On the other hand, if we 
assume that the observed motion is entirely due to orbital motion, 
then the emission source is located slightly off axis, 
trailing the black hole.  This could correspond to the location of 
a photoionization wake \citep{blo90} or a stream - disk interaction zone.  
The minimum disk radius estimate in the latter case would place the black hole 
at a position consistent with $q = (K_{\rm opt} \sin (-\phi_0)) /K_{\rm em}$. 
All these possibilities suggest $q \approx 0.36\pm0.05$, which is consistent 
with the range of mass ratios admitted by the light curve solutions 
\citep{gie86a}, but which is lower than the estimate derived from 
spectroscopic analysis of the supergiant by \citet{her95}. 

\placefigure{fig8}     % Figure 8 - orbital geometry diagram

If the moving emission component does originate in gas accreted near 
the black hole, then it is important to determine how the emission 
flux varies as the wind of the supergiant varies.  We illustrate in 
Figure~9 the reconstructed spectra of the supergiant and moving
spectral components for the same four groupings of emission 
equivalent width shown earlier in Figures 2 -- 5.   We see that 
there is a reasonably good correlation between the strength of the 
supergiant P~Cygni emission (related to the wind mass loss rate) and 
the strength of the moving emission component (presumably related 
to the gas accreted by the black hole).  We will return to the issue
of wind accretion in \S5. 

\placefigure{fig9}     % Figure 9 - reconstructions for each group 

%%%%%%%%%%%%%%%%%%%%%%%%%%%%%%%%%%%%%%%%%%%%%%%%%%%%%%%%%%%%%%%

\section{Correlated X-ray Variability}              % Section 4

The H$\alpha$ emission and associated stellar wind variations 
could be related to the X-ray flux variations, since the wind is 
probably the main source of material accreted by the black hole. 
Here we investigate the temporal variations through a 
comparison of the H$\alpha$ equivalent width with the 
X-ray flux observed with the RXTE/ASM 
instrument\footnote{http://xte.mit.edu/} \citep{lev96}.
The ASM records the X-ray flux in three bands (1.5 -- 3 keV, 
3 -- 5 keV, and 5 -- 12 keV), and we begin by plotting the daily average 
of the low energy band flux with the H$\alpha$ equivalent width in Figure~10. 
Our first runs occurred during the low/hard X-ray state 
when the soft X-ray flux was small.  However, the observations
made near HJD 2,451,896  
correspond to a time of X-ray flaring, while the DDO runs cover the 
recent, extended high/soft state of Cyg~X-1.  The general trend appears
to be an anti-correlation between $W_\lambda$(H$\alpha$) and the soft 
X-ray flux, and this same relationship is also found in some of the rapid variations. 
Figure~11 shows a detailed view of the co-variations observed during the 
third run in 1999, this time depicting the higher time resolution, 
dwell by dwell, X-ray fluxes.  The H$\alpha$ emission strength 
shows a general night-to-night trend, 
which was interrupted during one night (HJD 2,451,426.8)
when the emission dropped significantly.   We see that a small X-ray 
flare occurred at the same time as the H$\alpha$ decrease. 
Figure~12 shows an example of the reverse trend observed during 
our run with the best time coverage from 2000 December.  Here 
we see that the local maximum in H$\alpha$ emission strength 
occurred at nearly the same time as a well documented minimum in 
soft X-ray flux (HJD 2,451,898.6).  There appear to be no obvious 
time lags between the H$\alpha$ and X-ray events, although 
our time resolution is limited by the near diurnal sampling of the 
H$\alpha$ observations. 

\placefigure{fig10}     % Figure 10 - long term H-alpha and ASM data

\placefigure{fig11}     % Figure 11 - 1999B H-alpha and ASM data

\placefigure{fig12}     % Figure 12 - 2000B H-alpha and ASM data

We show the nature of the anti-correlation in all three X-ray bands
in Figure~13.  Here we have made a linear time interpolation in the 
dwell by dwell X-ray fluxes in order to make the best approximation 
of the flux values at the specific times of the H$\alpha$ observations. 
Note that the X-ray flux varies significantly on timescales shorter 
than dwell by dwell sampling rate (median interval of $\approx 0.8$ hours)
and that the X-ray and optical observations are often non-contemporaneous, 
so that our time-interpolation estimate of the X-ray flux may have large errors. 
The DDO observations made during the high/soft state are plotted
as diamonds while all the earlier measurements are shown as plus signs. 
These plots show that whenever we observed the H$\alpha$ emission 
to be stronger than $W_\lambda < -0.5$ \AA , the X-ray flux was 
relatively low (especially in the lower energy bands).   On the other hand, 
when the H$\alpha$ emission was weaker than this value, the X-ray 
flux appeared to vary over a wide range nearly independent of $W_\lambda$.   
We show a spectral hardness ratio in Figure 14 which confirms the 
impression that the X-ray spectrum is generally harder when the 
H$\alpha$ emission is stronger.   These trends were first observed 
by \citet{vol97} (based on data obtained during the 1996 high/soft state), 
and they also find confirmation in an important long term monitoring 
program of H$\alpha$ made at the Crimean Astrophysical Observatory 
(A. E. Tarasov 2002, private communication).  

\placefigure{fig13}     % Figure 13 - H-alpha and ASM data in 3 energy bands

\placefigure{fig14}     % Figure 14 - H-alpha and ASM data hardness ratio

%%%%%%%%%%%%%%%%%%%%%%%%%%%%%%%%%%%%%%%%%%%%%%%%%%%%%%%%%%%%%%%

\section{Discussion}                                % Section 5

Our results indicate that X-ray flux appears to be related to the 
state of the wind (as observed in the H$\alpha$ emission strength).  
When the H$\alpha$ emission is strong the X-ray 
flux is consistently low, but when the emission is 
relatively weak the X-ray flux can cover  
a wide range in values.   There are three processes that can  
potentially explain these observations: (1) X-ray photoionization 
of the wind leading to a decrease in H$\alpha$ emissivity;
(2) wind-related changes in column density towards the X-ray source; 
and (3) variations in wind strength that result in changes in 
accretion rate.  Here we argue that all three processes are required 
to help explain our results and that changes in the wind accretion rate 
may trigger the X-ray transitions in Cyg~X-1.  

The first possible explanation is that increased X-ray emission 
leads to heating of the wind and a corresponding decrease in H$\alpha$ 
emissivity \citep{bro99b}.  \citet{vlo01} show how the X-ray ionization 
of the wind in the low/hard state causes orbital variations in the 
ultraviolet P~Cygni lines formed in the wind (e.g., the Hatchett-McCray 
effect; \citet{hat77}).  The X-ray ionization effects become more severe
in the high/soft state, and \citet{wen99} describe how the wind absorption 
effects decrease significantly in such highly ionized conditions.
Hydrogen is expected to be nearly fully ionized in the base of the wind 
irrespective of the X-ray state, but if the gas is heated by an 
elevated soft X-ray flux, then the H$\alpha$ emissivity will drop 
($\propto T^{-1.2}$; \citet{ric98}).   

The best evidence for ionization-related variability 
is the reduction in H$\alpha$ equivalent width we observed during 
the mini-flare of X-ray flux around HJD 2,451,426.8 (Fig.~11). 
The increase in soft X-ray flux and reduction in H$\alpha$ occurred
more-or-less simultaneously (within the time resolution of the data),
as expected for the small light-travel time between the X-ray source 
and the facing hemisphere of the supergiant ($\approx 54$ s).  
On the other hand, the characteristic wind crossing time between components 
is approximately 10 hours, and our data, although sparse in time coverage, 
does not indicate any time lag of this order between the 
H$\alpha$ and X-ray variations.   This event occurred near 
orbital phase $\phi=0.77$, and the two spectra we obtained then show H$\alpha$ 
profiles with unusually deep absorption cores (see Fig.~4) compared 
to observations on the preceding and following nights.  At this 
orientation the wind between the stars was moving tangentially 
to our line of sight and the emission from this region, if present,  
would have filled in the line core.  However, if this inner region was 
significantly photoionized by the X-ray mini-flare, then the 
residual emission in the line core would have vanished to produce the deeper
absorption core.  Thus, the timing and line properties of this 
event suggest it was caused by X-ray photoionization.

On the other hand, there are several lines of evidence that 
X-ray photoionization is not the dominant cause of the H$\alpha$
variations.  First, we would expect the anti-correlation between 
H$\alpha$ emission and X-ray flux to be better defined than exhibited 
in Figure~13 if X-ray photoionization drove the emission variations
(although the scatter may be partially the result of the poor time resolution 
and poor overlap of the H$\alpha$ and X-ray observations).  There are several 
examples (especially during the X-ray low/hard state) where we observed 
large excursions in the H$\alpha$ emission while the X-ray 
flux was essentially constant (Fig.~10).  
Second, if X-ray photoionization significantly altered the 
H$\alpha$ emissivity of the gas above the hemisphere facing the X-ray source, 
then we should observe changes in the shape of the supergiant's P~Cygni 
component with orbital phase.  The variations would be largest 
in the high/soft X-ray state and would result in decreased blue 
absorption near $\phi = 0.5$ and decreased red emission near $\phi = 0.0$.
The spectra observed in the high/soft X-ray state (Fig.~5) 
show that the predicted changes may be present but are relatively 
minor in nature.   Third, we find evidence of 
large scale variations in the H$\alpha$ emission 
that forms in the X-ray shadow above the hemisphere facing away 
from the X-ray source where no X-ray photoionization should occur. 
This region in the wind is best isolated in the radial velocity 
distribution of the emission near orbital phase $\phi=0.5$  
when this backside outflow is the dominant contributor to 
the red-shifted peak of the P~Cygni profile.   However, 
given the probable low orbital inclination, this part of the 
profile also includes some contributions from the X-ray 
illuminated hemisphere.  We show in 
Figure~15 the red peak emission height for spectra obtained 
near this phase plotted against X-ray flux.   These diagrams 
show that the strength of the emission component 
from the X-ray shadow region is generally independent of 
the X-ray flux level in both the low/hard ({\it plus signs}) and 
high/soft ({\it diamonds}) X-ray states,  
as expected for an emitting region sheltered by the supergiant.
In contrast, we show in Figure~16 the peak intensity of the 
blue emission peak isolated near orbital phase $\phi = 0.25$ 
that displays evidence of an anti-correlation between 
H$\alpha$ emission and X-ray flux as expected for irradiated 
gas close to the black hole (\S3).  
The large range in the strength of the H$\alpha$ emission from the 
X-ray shadow region suggests that variations are caused 
by factors other than changes in X-ray photoionization. 
Since the emission strength is very sensitive to gas density ($\propto n^2$), 
we conclude that the main source of the H$\alpha$ emission variations 
is the fluctuation in basal wind density rather than ionization state. 

\placefigure{fig15}     % Figure 15 - H-alpha red peak near 0.5 and ASM data 
                                                                                
\placefigure{fig16}     % Figure 16 - H-alpha blue peak near 0.25 and ASM data 
                                                                                
If the H$\alpha$ variations primarily reflect changes in wind density close to the supergiant, 
then it is possible that the X-ray flux is partially modulated by the changing 
absorption in the stellar wind (particularly important for soft X-rays).   
We doubt that wind absorption plays a major role in explaining 
the general anti-correlation between H$\alpha$ and X-ray emission
because the scatter in their temporal variations (Fig.~13) is much larger 
than we would expect for a direct cause-and-effect relationship. 
Furthermore, the amplitude of the wind density fluctuations implied by the 
H$\alpha$ changes is too small to explain the range in X-ray variability. 
\citet{wen99} show that the X-ray light curve can be explained 
by the changing column density of the line of sight to the black hole 
as we peer through different portions of the supergiant's wind 
(strongest absorption at supergiant inferior conjunction, $\phi=0.0$). 
Their ASM 1.5 -- 3 keV orbital light curve for the low/hard state shows a 
25\% decrease at $\phi=0.0$ relative to $\phi=0.5$.  \citet{bal00} also 
studied the variation in column density with orbital phase, and they suggest 
that the orbital modulation in X-ray flux corresponds to a fluctuation of 
$6 - 10\times$ in column density (depending on the assumed ionization of the wind). 
However, the observed variations in the emission strength generally indicate column 
density changes of a factor of $<2$ (see below).   Wind column density changes of 
this order are too small to account for the large changes in X-ray flux. 
Thus, wind absorption variations are insufficient to explain the general 
anti-correlation between the H$\alpha$ emission strength and X-ray flux. 
Nevertheless, we might expect the wind absorption processes to appear 
most prominent near inferior conjunction of the supergiant when 
the X-ray light curve attains a minimum and a peak occurs in the 
frequency of rapid X-ray dips \citep{wen99,bal00,fen02}.  It is 
noteworthy in this regard that the case of the simultaneous H$\alpha$ maximum and 
X-ray flux minimum shown in Figure~12 did indeed occur near this phase 
(at $\phi = 0.02$).  

If we ignore the effects of X-ray ionization 
on the H$\alpha$ emissivity, then we can make an approximate estimate 
of how the wind mass loss rate and density vary as a function 
of the H$\alpha$ emission strength \citep{pul96}.  
\citet{her95} show predictions of how the H$\alpha$ profile will vary as a function of 
mass loss rate (see their Fig.~4), and we measured the equivalent widths 
of their profiles to calibrate the mass loss rate as a function of 
H$\alpha$ equivalent width (prorated to their final estimate of 
mass loss rate for the time of their observations, $3.0\times 10^{-6}$ 
$M_\odot$~y$^{-1}$).   The functional fit in this case is 
\begin{equation}
-\dot{M} = (1.85 -1.01~W_\lambda - 0.04~W_\lambda^2)~\times 10^{-6}~M_\odot~{\rm y}^{-1}.
\end{equation}
We grouped our equivalent width data according to the X-ray 
state at the time of observation (setting aside the results 
obtained near HJD 2,451,896 that may correspond to a ``failed 
transition''), and the mean equivalent width for each group
yields mass loss rates of $(2.57\pm 0.05)\times 10^{-6}$
and $(2.00\pm 0.03)\times 10^{-6}$ $M_\odot$~y$^{-1}$ 
for the low/hard and high/soft states, respectively
(the quoted errors are based on the standard deviation of the mean
and do not include the larger errors associated with the calibration of the 
$W_\lambda - \dot{M}$ relationship and with our neglect of the 
multiple component nature of the H$\alpha$ emission).  
This suggests that the mass loss rate is $\approx 22\%$ lower 
during the rare high/soft state compared to the more common low/hard state.
Since this calculation ignores the X-ray photoionization 
that may be more important in the high/soft state and since
photoionization will decrease H$\alpha$ emission strength, 
our estimate for the mass loss rate during the
high/soft state is best regarded as a lower limit.  
Nevertheless, the general decrease in wind strength during 
the high/soft state is also observed in the emission from 
the X-ray shadow region (Fig.~15) that should be relatively 
free from the effects of photoionization.  We find that the 
mean residual emission intensity of the red peak near 
phase $\phi = 0.5$ is $0.139\pm 0.008$ and $0.114\pm 0.006$ 
for the low/hard and high/soft states, respectively.  
\citet{wen99} also found that the wind mass loss rate is lower 
in the high/soft state based upon their analysis of the 
X-ray orbital light curve.  

Taken at face value, our results present a paradox: the X-ray flux 
decreases when the wind mass loss rate increases.    We suggest that 
the resolution of this quandary lies in how the wind velocity changes 
with wind ionization state (originally proposed by \citet{ho87} and
developed in more physical terms by \citet{ste91}).   
When the wind mass loss rate is high, 
the wind density is also proportionally high, and therefore the 
X-ray ionization effects on the wind are confined to the region close 
to the black hole since the size of the surrounding ionized region 
depends on wind density as $n^{-1}$ \citep{ste91,vlo01}. 
Most of the wind volume surrounding the star will contain the 
many important ions that propel the radiative driving
of the wind, so that it reaches a terminal velocity of approximately 
2100 km~s$^{-1}$ \citep{her95}.   However, the wind flow close to the 
black hole will become ionized, and these advanced ionization 
states will generally have transitions at frequencies much higher 
than the peak of the stellar flux distribution.  Consequently  
there are fewer absorbing transitions at frequencies where 
the stellar flux is concentrated and radiative driving of the 
wind becomes less effective.  Stellar wind gas leaving the 
part of the supergiant facing the X-ray source will 
be accelerated by radiative driving, reaching a significant 
fraction of the terminal velocity before crossing  
the distant ionization boundary \citep{ste91}.  Models of wind accretion by 
the black hole suggest that the mass accretion rate is proportional 
to $\dot{M}/ v^4$ \citep{bon44,lam76,ste91}, and since the flow velocity $v$ is 
relatively large, only modest mass accretion occurs.   It is easy to imagine 
that some of the outflow from the supergiant to the black hole bypasses 
the black hole altogether, and absorption by the gas beyond the black hole 
could explain the presence of a secondary minimum in the low/hard 
state X-ray light curve at orbital phase $\phi=0.5$ \citep{kar01}.  

However, when the mass loss rate drops, the X-ray ionization zone  
will become larger because of the reduced wind density. 
The dominant ions that provide the resonant transitions that in turn drive 
the stellar wind will disappear with increased ionization, and the 
wind outflow will experience only a modest acceleration and reach 
a speed of just a few hundred km~s$^{-1}$ (before the flow dynamics become
dominated by the gravitational acceleration of the black hole).   
This altered outflow will be denser 
and slower in the vicinity of the black hole, and since the 
accretion rate varies as $\dot{M}/ v^4$, the overall accretion rate 
will increase significantly because of the slower outflow (and 
despite a real decline in $\dot{M}$).   

If this basic scenario is correct, then transitions from the low/hard 
to the high/soft state are triggered when the supergiant undergoes an episode of 
reduced mass loss.   \citet{ste91} demonstrates that the decrease 
in the wind force multiplier occurs rather suddenly once a specific 
gas density is reached, and we would argue that this is the reason why 
the transitions are relatively fast and the X-ray states are bimodal.  
Once the transition occurs, the system tends 
to remain in the high/soft state because the increased accretion rate 
and associated larger X-ray fluxes make it easier to keep the wind 
in the highly ionized state (even with modest increases in mass loss rate).
The line of sight to the black hole then passes through mainly ionized 
gas all around the orbit, and the wind absorption of the X-ray flux 
decreases so much that the X-ray light curve modulation vanishes  
\citep{wen99}.  
The increased mass accretion in this state causes
the optically thick, geometrically thin (Keplerian) accretion
disk to extend further inwards towards the black hole,
producing more soft X-rays, while the coronal (ADAF or
sub-Keplerian) region, which produces the bulk of the hard
X-rays, becomes smaller, and thus the X-ray spectrum softens
\citep{ebi96,esi98,bro99b}.
The enhanced soft X-ray flux heats the accreting gas, 
which reduces the H$\alpha$ emission from the accretion flow (Fig.~9, 16).  
It is only once the supergiant can maintain a strong mass outflow 
that the wind becomes dense enough to reduce the ionization effects 
and to create the high speeds which then lower the mass 
accretion rate and force the system back to the
low/hard state.   The long term variations in the mass loss rates 
of massive supergiant stars are not well documented, but 
there is circumstantial evidence of significant 
variations on time scales of years \citep{ebb82,mar02}.   Thus, we
suggest that the high/soft states that occur every 5 years or so in Cyg~X-1 
correspond to quasi-cyclic minima in the mass loss rate of the supergiant. 

%%%%%%%%%%%%%%%%%%%%%%%%%%%%%%%%%%%%%%%%%%%%%%%%%%%%%%%%%%%%%%%

\acknowledgements

We are grateful to Dr.\ Anatoly Tarasov for sharing his 
H$\alpha$ observational results with us in advance of publication, 
and we thank Dr.\ Ian Stevens for helpful comments on our work.  
We also thank the KPNO staff, and in particular Diane Harmer
and Daryl Willmarth, for their assistance in making these
observations with the KPNO Coude Feed Telescope.
We are grateful to the staff of McDonald Observatory and
Tom Montemayor for their help at the 2.1-m telescope.
This research has made use of results provided by the
RXTE/ASM teams at MIT and at the RXTE SOF and GOF at
the NASA/Goddard Space Flight Center.
Support for this work was provided by NASA through grant numbers
GO-8308 and GO-9449 from the Space Telescope Science Institute, which is
operated by the Association of Universities for Research in
Astronomy, Inc., under NASA contract NAS5-26555.
Institutional support has been provided from the GSU College
of Arts and Sciences and from the Research Program Enhancement
fund of the Board of Regents of the University System of Georgia,
administered through the GSU Office of the Vice President for Research.
P.\ J.\ Wiita is grateful for hospitality from the Princeton University 
Department of Astrophysical Sciences.
C.\ T.\ Bolton's research is partially supported
by a Discovery Grant from the Natural Sciences and Engineering Research
Council of Canada.  The University of Toronto funds the operation of the
David Dunlap Observatory.
We gratefully acknowledge all this support.

%%%%%%%%%%%%%%%%%%%%%%%%%%%%%%%%%%%%%%%%%%%%%%%%%%%%%%%%%%%%%%%

% References

%%%%%%%%%%%%%%%%%%%%%%%%%%%%%%%%%%%%%%%%%%%%%%%%%%%%%%%%%%%%%%%

% Figures

\clearpage

\begin{figure}
\caption{Radial velocity curve ({\it solid line}) derived 
from measurements of \ion{He}{1} $\lambda 6678$ ({\it filled circles}).
Phase 0.0 is inferior conjunction of the supergiant.}
\label{fig1}
\end{figure}

\begin{figure}
\caption{
{\it Upper frame}: Strong emission H$\alpha$ profiles 
($W_\lambda < -0.9$ \AA ) as a function of heliocentric radial velocity.
The profiles are arranged in order of increasing
orbital phase and each is placed in the $y$ ordinate so that
the continuum equals the phase of observation. 
{\it Lower frame}:
A gray-scale representation of the profiles above.
The profile at each phase is calculated by a linear interpolation between
the closest observed phases (marked by arrows on the right hand side). 
The gray intensity is scaled between spectral intensity 0.94 (black) 
and 1.15 (white) here and in Figs.\ 3 -- 5. 
The first and last 20\% of the orbit have been
reproduced at the bottom and top of the image, respectively, 
to improve the sense of phase continuity.
The white lines show the radial velocity curves of the 
supergiant and anti-phase moving component.}
\label{fig2}
\end{figure}

\begin{figure}
\caption{Moderate emission H$\alpha$ profiles ($-0.9 < W_\lambda < -0.4$ \AA )
in the same format as Fig.~2.}
\label{fig3}
\end{figure}

\begin{figure}
\caption{Weak emission H$\alpha$ profiles ($W_\lambda > -0.4$ \AA )
in the same format as Fig.~2.  The spectrum obtained just 3 hours prior to 
the {\it Chandra} observation \citep{sch02} is shown as a thick line.}
\label{fig4}
\end{figure}

\begin{figure}
\caption{H$\alpha$ profiles obtained during the 2001--2002 X-ray high/soft state
(in the same format as Fig.~2).}
\label{fig5}
\end{figure}

\begin{figure}
\caption{Root-mean-square residuals from the fit of a two
component tomographic reconstruction.  The first component is 
assumed to move with the orbital motion of the supergiant star 
while the second component is parameterized by an 
orbital phase of maximum radial velocity, $\phi_0$, 
and a semiamplitude, $K_{\rm em}$.  The smallest residuals are shown
in white, the largest in black.}
\label{fig6}
\end{figure}

\begin{figure}
\caption{Reconstructed spectra of the component moving 
with the supergiant ({\it left panel}) and the 
component moving with parameters than minimize the 
rms residuals ({\it right panel}).  The three parameter 
sets illustrated are 
(A) $\phi_0=0.79$ and $K_{\rm em}=218$ km~s$^{-1}$ ({\it solid line}), 
(B) $\phi_0=0.81$ and $K_{\rm em}=56$ km~s$^{-1}$ ({\it dotted line}), and
(C) $\phi_0=0.23$ and $K_{\rm em}=185$ km~s$^{-1}$ ({\it dashed line}). 
The continua levels are offset for clarity by $-0.15$ and $-0.30$ for cases
B and C, respectively.}
\label{fig7}
\end{figure}

\begin{figure}
\caption{
A diagram of the system geometry as viewed from
above the orbital plane.  The figure of the supergiant was
calculated (ignoring radiation pressure) for a mass ratio, $q=0.36$,
and a primary fill-out factor, $\rho = 0.97$ (see Gies \& Bolton 1986a).
Orbital phases are listed around the periphery.
The arrow at the stellar origin indicates the projected orbital velocity
while the center-of-mass position is indicated by CM.
The projected velocity vectors at positions near the secondary show the
observed semiamplitude $K_{\rm em}$ of the anti-phase moving component and
its derived orbital and flow components for two cases discussed
in the text.}
\label{fig8}
\end{figure}

\begin{figure}
\caption{Reconstructed spectra of the component moving 
with the supergiant ({\it left panel}) and the 
anti-phase moving component ({\it right panel}).  
The different plots correspond to samples for times of 
(from top to bottom) strong emission (Fig.~2), moderate emission (Fig.~3), 
weak emission (Fig.~4), and the X-ray high/soft state (Fig.~5).
The continua levels are each offset by 0.10.}
\label{fig9}
\end{figure}

\begin{figure}
\caption{A comparison of the H$\alpha$ equivalent width with 
the daily average RXTE/ASM count rate in the 1.5 -- 3 keV range
over the entire course of our observations. The error bar in the upper
panel gives the typical measurement error for $W_\lambda ({\rm H}\alpha)$.}
\label{fig10}
\end{figure}

\begin{figure}
\caption{A detailed comparison of the H$\alpha$ equivalent width with 
the dwell RXTE/ASM count rate in the 1.5 -- 3 keV range
showing an emission weakening episode near an X-ray mini-flare
(HJD 2,451,426.8).}
\label{fig11}
\end{figure}

\begin{figure}
\caption{A detailed comparison of the H$\alpha$ equivalent width with 
the dwell RXTE/ASM count rate in the 1.5 -- 3 keV range
showing an emission strengthening -- X-ray weakening event
near HJD 2,451,898.7.}
\label{fig12}
\end{figure}

\begin{figure}
\caption{The H$\alpha$ equivalent widths versus
the time-interpolated dwell RXTE/ASM count rates in 
the 1.5 -- 3 keV ({\it top}), 
3 -- 5 keV ({\it middle}), and 
5 -- 12 keV ranges ({\it bottom}). 
Diamonds correspond to the high/soft state data while pluses
correspond to the low/hard state data.}
\label{fig13}
\end{figure}

\begin{figure}
\caption{The H$\alpha$ equivalent widths versus the
ratio of the 5 -- 12 keV to 1.5 -- 3 keV band
time-interpolated dwell RXTE/ASM count rates. 
Diamonds correspond to the high/soft state data while pluses
correspond to the low/hard state data.}
\label{fig14}
\end{figure}

\begin{figure}
\caption{The residual intensities of the red peak of 
the H$\alpha$ P~Cygni profile for spectra obtained 
near orbital phase $\phi=0.5$ versus
the time-interpolated dwell RXTE/ASM count rates
(in the same format as Fig.~13).  This part of the H$\alpha$
emission forms in a region where X-rays are mostly blocked 
by the supergiant so that photoionization effects are
minimized.}
\label{fig15}
\end{figure}

\begin{figure}
\caption{The residual intensities of the blue peak of 
the H$\alpha$ profile for spectra obtained 
near orbital phase $\phi=0.25$ versus
the time-interpolated dwell RXTE/ASM count rates
(in the same format as Fig.~13).  This part of the H$\alpha$
emission forms in a region close to the black hole where 
X-ray photoionization does appear to affect the emission strength.
The outlying points probably result from time interpolation 
of the X-ray flux in intervals when the optical and X-ray 
observations are non-contemporaneous.}
\label{fig16}
\end{figure}

%%%%%%%%%%%%%%%%%%%%%%%%%%%%%%%%%%%%%%%%%%%%%%%%%%%%%%%%%%%%%%%

% Tables

\clearpage

% Table 1
\begin{deluxetable}{lccccl}
\tabletypesize{\scriptsize}
\tablewidth{0pt}
\tablenum{1}
\tablecaption{Radial Velocity and Equivalent Width Measurements \label{tab1}}
\tablehead{
\colhead{HJD}             &
\colhead{Orbital}         &
\colhead{$V_r$}           &
\colhead{$(O-C)$}       &
\colhead{$W_\lambda$}     &
\colhead{Observatory/Telescope/}      \\
\colhead{(-2,450,000)}    &
\colhead{Phase}           &
\colhead{(km s$^{-1}$)}   &
\colhead{(km s$^{-1}$)}   &
\colhead{(\AA )}          &
\colhead{Grating/CCD} }
\scriptsize
\startdata
 1053.7473 \dotfill &  0.157 &
\phn\phs     $  52.5$ &\phn     $  -3.5$ &
     $-0.73$ &
 KPNO/0.9m/B/F3KB         \\
 1053.7687 \dotfill &  0.160 &
\phn\phs     $  58.1$ &\phn\phs $   1.1$ &
     $-0.83$ &
 KPNO/0.9m/B/F3KB         \\
 1055.7247 \dotfill &  0.510 &
\phn\phn     $  -4.5$ &\phn\phs $   7.1$ &
     $-0.91$ &
 KPNO/0.9m/B/F3KB         \\
 1055.7830 \dotfill &  0.520 &
\phn\phn     $  -9.1$ &\phn\phs $   7.4$ &
     $-0.91$ &
 KPNO/0.9m/B/F3KB         \\
 1056.7695 \dotfill &  0.696 &
\phn         $ -79.0$ &\phn     $  -0.7$ &
     $-1.57$ &
 KPNO/0.9m/B/F3KB         \\
 1056.7908 \dotfill &  0.700 &
\phn         $ -81.3$ &\phn     $  -2.4$ &
     $-1.71$ &
 KPNO/0.9m/B/F3KB         \\
 1057.7653 \dotfill &  0.874 &
\phn         $ -68.2$ &\phn     $  -7.5$ &
     $-1.53$ &
 KPNO/0.9m/B/F3KB         \\
 1057.7864 \dotfill &  0.878 &
\phn         $ -66.0$ &\phn     $  -6.6$ &
     $-1.66$ &
 KPNO/0.9m/B/F3KB         \\
 1058.7556 \dotfill &  0.051 &
\phn\phs     $  23.2$ &\phn\phs $   6.4$ &
     $-0.71$ &
 KPNO/0.9m/B/F3KB         \\
 1058.7766 \dotfill &  0.055 &
\phn\phs     $  25.6$ &\phn\phs $   7.1$ &
     $-0.90$ &
 KPNO/0.9m/B/F3KB         \\
 1061.7516 \dotfill &  0.586 &
\phn         $ -49.2$ &\phn     $  -3.3$ &
     $-0.73$ &
 KPNO/0.9m/B/F3KB         \\
 1061.7733 \dotfill &  0.590 &
\phn         $ -48.5$ &\phn     $  -1.1$ &
     $-0.74$ &
 KPNO/0.9m/B/F3KB         \\
 1062.8785 \dotfill &  0.787 &
\phn         $ -86.0$ &\phn     $  -5.5$ &
     $-0.96$ &
 KPNO/0.9m/B/F3KB         \\
 1063.8615 \dotfill &  0.963 &
\phn         $ -28.0$ &\phn     $  -3.5$ &
     $-1.22$ &
 KPNO/0.9m/B/F3KB         \\
 1065.7631 \dotfill &  0.302 &
\phn\phs     $  57.7$ &\phn     $  -6.8$ &
     $-0.56$ &
 KPNO/0.9m/B/F3KB         \\
 1065.7846 \dotfill &  0.306 &
\phn\phs     $  55.2$ &\phn     $  -8.7$ &
     $-0.75$ &
 KPNO/0.9m/B/F3KB         \\
 1066.7063 \dotfill &  0.471 &
\phn\phn     $  -0.9$ &\phn     $  -7.7$ &
     $-0.72$ &
 KPNO/0.9m/B/F3KB         \\
 1066.7276 \dotfill &  0.475 &
\phn\phn     $  -3.0$ &\phn     $  -8.0$ &
     $-0.92$ &
 KPNO/0.9m/B/F3KB         \\
 1354.8761 \dotfill &  0.931 &
\phn         $ -41.2$ &\phn     $  -2.6$ &
     $-0.39$ &
 KPNO/0.9m/RC181/TI5      \\
 1355.8228 \dotfill &  0.100 &
\phn\phs     $  33.5$ &\phn     $  -4.1$ &
     $-0.89$ &
 KPNO/0.9m/RC181/F3KB     \\
 1356.7938 \dotfill &  0.274 &
\phn\phs     $  75.5$ &\phn\phs $   7.7$ &
     $-0.49$ &
 KPNO/0.9m/RC181/F3KB     \\
 1357.6648 \dotfill &  0.429 &
\phn\phs     $  17.8$ &\phn     $  -7.7$ &
     $-1.00$ &
 KPNO/0.9m/RC181/F3KB     \\
 1358.8837 \dotfill &  0.647 &
\phn         $ -67.2$ &\phn\phs $   0.0$ &
     $-1.16$ &
 KPNO/0.9m/RC181/F3KB     \\
 1359.7987 \dotfill &  0.810 &
\phn         $ -87.7$ &         $ -10.5$ &
     $-1.00$ &
 KPNO/0.9m/RC181/F3KB     \\
 1360.7959 \dotfill &  0.988 &
\phn         $ -16.0$ &\phn     $  -3.5$ &
     $-0.92$ &
 KPNO/0.9m/RC181/F3KB     \\
 1361.6782 \dotfill &  0.146 &
\phn\phs     $  48.6$ &\phn     $  -4.4$ &
     $-1.35$ &
 KPNO/0.9m/RC181/F3KB     \\
 1362.7495 \dotfill &  0.337 &
\phn\phs     $  57.0$ &\phn     $  -0.5$ &
     $-0.80$ &
 KPNO/0.9m/RC181/F3KB     \\
 1363.6586 \dotfill &  0.500 &
\phn\phn\phs $   0.3$ &\phn\phs $   7.1$ &
     $-1.39$ &
 KPNO/0.9m/RC181/F3KB     \\
 1394.8165 \dotfill &  0.064 &
\phn\phs     $  15.1$ &\phn     $  -7.4$ &
     $-0.83$ &
 McD/2.1m/Echelle/RA2     \\
 1396.8232 \dotfill &  0.422 &
\phn\phs     $  23.2$ &\phn     $  -5.4$ &
     $-0.96$ &
 McD/2.1m/Echelle/RA2     \\
 1398.8853 \dotfill &  0.790 &
\phn         $ -80.1$ &\phn\phs $   0.0$ &
     $-0.34$ &
 McD/2.1m/Echelle/RA2     \\
 1399.8281 \dotfill &  0.959 &
\phn         $ -34.0$ &\phn     $  -7.6$ &
\phs $ 0.00$ &
 McD/2.1m/Echelle/RA2     \\
 1420.8900 \dotfill &  0.720 &
\phn         $ -73.1$ &\phn\phs $   8.1$ &
     $-0.11$ &
 KPNO/0.9m/RC181/F3KB     \\
 1421.7939 \dotfill &  0.881 &
\phn         $ -62.0$ &\phn     $  -3.7$ &
     $-0.30$ &
 KPNO/0.9m/RC181/F3KB     \\
 1421.8155 \dotfill &  0.885 &
\phn         $ -62.8$ &\phn     $  -5.9$ &
     $-0.30$ &
 KPNO/0.9m/RC181/F3KB     \\
 1423.8200 \dotfill &  0.243 &
\phn\phs     $  75.3$ &\phn\phs $   6.8$ &
     $-0.37$ &
 KPNO/0.9m/RC181/F3KB     \\
 1425.8079 \dotfill &  0.598 &
\phn         $ -41.4$ &\phn\phs $   9.2$ &
     $-0.46$ &
 KPNO/0.9m/RC181/F3KB     \\
 1426.7688 \dotfill &  0.770 &
\phn         $ -76.4$ &\phn\phs $   5.6$ &
     $-0.17$ &
 KPNO/0.9m/RC181/F3KB     \\
 1426.7920 \dotfill &  0.774 &
\phn         $ -77.0$ &\phn\phs $   4.7$ &
     $-0.12$ &
 KPNO/0.9m/RC181/F3KB     \\
 1427.7710 \dotfill &  0.949 &
\phn         $ -27.7$ &\phn\phs $   3.2$ &
     $-0.60$ &
 KPNO/0.9m/RC181/F3KB     \\
 1428.7417 \dotfill &  0.122 &
\phn\phs     $  50.4$ &\phn\phs $   5.0$ &
     $-0.50$ &
 KPNO/0.9m/RC181/F3KB     \\
 1429.7366 \dotfill &  0.300 &
\phn\phs     $  66.4$ &\phn\phs $   1.4$ &
     $-0.44$ &
 KPNO/0.9m/RC181/F3KB     \\
 1429.7576 \dotfill &  0.303 &
\phn\phs     $  66.3$ &\phn\phs $   1.9$ &
     $-0.45$ &
 KPNO/0.9m/RC181/F3KB     \\
 1464.6959 \dotfill &  0.543 &
\phn         $ -30.3$ &\phn     $  -3.4$ &
     $-0.70$ &
 KPNO/0.9m/RC181/F3KB     \\
 1465.7047 \dotfill &  0.723 &
\phn         $ -79.9$ &\phn\phs $   1.5$ &
     $-0.67$ &
 KPNO/0.9m/RC181/F3KB     \\
 1466.6863 \dotfill &  0.898 &
\phn         $ -53.0$ &\phn     $  -0.9$ &
     $-0.29$ &
 KPNO/0.9m/RC181/F3KB     \\
 1467.6950 \dotfill &  0.078 &
\phn\phs     $  25.5$ &\phn     $  -3.2$ &
     $-0.25$ &
 KPNO/0.9m/RC181/F3KB     \\
 1467.7170 \dotfill &  0.082 &
\phn\phs     $  29.3$ &\phn     $  -1.0$ &
     $-0.40$ &
 KPNO/0.9m/RC181/F3KB     \\
 1469.6884 \dotfill &  0.434 &
\phn\phs     $  13.3$ &         $ -10.1$ &
     $-0.70$ &
 KPNO/0.9m/RC181/F3KB     \\
 1469.7095 \dotfill &  0.438 &
\phn\phs     $  15.0$ &\phn     $  -6.8$ &
     $-0.62$ &
 KPNO/0.9m/RC181/F3KB     \\
 1491.6827 \dotfill &  0.362 &
\phn\phs     $  43.4$ &\phn     $  -7.3$ &
     $-0.89$ &
 KPNO/0.9m/RC181/F3KB     \\
 1492.6604 \dotfill &  0.536 &
\phn         $ -25.4$ &\phn     $  -1.3$ &
     $-0.68$ &
 KPNO/0.9m/RC181/F3KB     \\
 1493.6465 \dotfill &  0.712 &
\phn         $ -86.4$ &\phn     $  -6.0$ &
     $-0.97$ &
 KPNO/0.9m/RC181/F3KB     \\
 1494.6562 \dotfill &  0.893 &
\phn         $ -60.8$ &\phn     $  -6.7$ &
     $-1.31$ &
 KPNO/0.9m/RC181/F3KB     \\
 1495.6573 \dotfill &  0.072 &
\phn\phs     $  23.9$ &\phn     $  -2.0$ &
     $-1.19$ &
 KPNO/0.9m/RC181/F3KB     \\
 1496.6541 \dotfill &  0.250 &
\phn\phs     $  74.2$ &\phn\phs $   5.6$ &
     $-0.91$ &
 KPNO/0.9m/RC181/F3KB     \\
 1497.6574 \dotfill &  0.429 &
\phn\phs     $  27.8$ &\phn\phs $   2.0$ &
     $-1.01$ &
 KPNO/0.9m/RC181/F3KB     \\
 1817.6317 \dotfill &  0.569 &
\phn         $ -42.0$ &\phn     $  -3.4$ &
     $-0.38$ &
 KPNO/0.9m/B/F3KB         \\
 1818.6805 \dotfill &  0.756 &
\phn         $ -86.9$ &\phn     $  -4.4$ &
     $-0.28$ &
 KPNO/0.9m/B/F3KB         \\
 1819.6634 \dotfill &  0.932 &
\phn         $ -37.0$ &\phn\phs $   1.5$ &
     $-0.73$ &
 KPNO/0.9m/B/F3KB         \\
 1820.6465 \dotfill &  0.107 &
\phn\phs     $  42.3$ &\phn\phs $   2.2$ &
     $-0.43$ &
 KPNO/0.9m/B/F3KB         \\
 1821.6565 \dotfill &  0.287 &
\phn\phs     $  68.2$ &\phn\phs $   1.7$ &
     $-0.47$ &
 KPNO/0.9m/B/F3KB         \\
 1822.6660 \dotfill &  0.468 &
\phn\phn\phs $   0.8$ &\phn     $  -7.4$ &
     $-0.22$ &
 KPNO/0.9m/B/F3KB         \\
 1823.6554 \dotfill &  0.644 &
\phn         $ -63.0$ &\phn\phs $   3.5$ &
     $-0.78$ &
 KPNO/0.9m/B/F3KB         \\
 1824.6356 \dotfill &  0.819 &
\phn         $ -70.5$ &\phn\phs $   4.9$ &
     $-0.56$ &
 KPNO/0.9m/B/F3KB         \\
 1890.5663 \dotfill &  0.593 &
\phn         $ -45.2$ &\phn\phs $   3.5$ &
     $-0.15$ &
 KPNO/0.9m/B/F3KB         \\
 1890.5879 \dotfill &  0.597 &
\phn         $ -48.5$ &\phn\phs $   1.7$ &
     $-0.05$ &
 KPNO/0.9m/B/F3KB         \\
 1891.5880 \dotfill &  0.776 &
\phn         $ -83.9$ &\phn     $  -2.3$ &
\phs $ 0.23$ &
 KPNO/0.9m/B/F3KB         \\
 1892.5719 \dotfill &  0.951 &
\phn         $ -27.9$ &\phn\phs $   1.8$ &
\phs $ 0.30$ &
 KPNO/0.9m/B/F3KB         \\
 1892.5933 \dotfill &  0.955 &
\phn         $ -24.6$ &\phn\phs $   3.4$ &
\phs $ 0.26$ &
 KPNO/0.9m/B/F3KB         \\
 1893.6007 \dotfill &  0.135 &
\phn\phs     $  46.7$ &\phn     $  -3.0$ &
\phs $ 0.05$ &
 KPNO/0.9m/B/F3KB         \\
 1893.6217 \dotfill &  0.139 &
\phn\phs     $  48.4$ &\phn     $  -2.5$ &
     $-0.29$ &
 KPNO/0.9m/B/F3KB         \\
 1894.5918 \dotfill &  0.312 &
\phn\phs     $  72.4$ &\phn\phs $   9.5$ &
     $-0.21$ &
 KPNO/0.9m/B/F3KB         \\
 1894.6129 \dotfill &  0.316 &
\phn\phs     $  73.8$ &\phs     $  11.6$ &
     $-0.24$ &
 KPNO/0.9m/B/F3KB         \\
 1895.5601 \dotfill &  0.485 &
\phn\phn\phs $   3.5$ &\phn\phs $   3.3$ &
     $-0.34$ &
 KPNO/0.9m/B/F3KB         \\
 1895.5811 \dotfill &  0.489 &
\phn\phn\phs $   2.4$ &\phn\phs $   4.0$ &
     $-0.37$ &
 KPNO/0.9m/B/F3KB         \\
 1896.5645 \dotfill &  0.664 &
\phn         $ -69.0$ &\phn\phs $   2.8$ &
     $-0.09$ &
 KPNO/0.9m/B/F3KB         \\
 1896.5856 \dotfill &  0.668 &
\phn         $ -71.6$ &\phn\phs $   1.1$ &
\phs $ 0.06$ &
 KPNO/0.9m/B/F3KB         \\
 1897.5610 \dotfill &  0.842 &
\phn         $ -65.6$ &\phn\phs $   4.6$ &
     $-0.28$ &
 KPNO/0.9m/B/F3KB         \\
 1897.5822 \dotfill &  0.846 &
\phn         $ -65.0$ &\phn\phs $   4.2$ &
     $-0.35$ &
 KPNO/0.9m/B/F3KB         \\
 1898.5674 \dotfill &  0.022 &
\phn\phn\phs $   7.8$ &\phn\phs $   4.4$ &
     $-0.83$ &
 KPNO/0.9m/B/F3KB         \\
 1898.5886 \dotfill &  0.026 &
\phn\phn\phs $   9.5$ &\phn\phs $   4.3$ &
     $-0.87$ &
 KPNO/0.9m/B/F3KB         \\
 1899.5713 \dotfill &  0.201 &
\phn\phs     $  63.6$ &\phn     $  -1.5$ &
     $-0.14$ &
 KPNO/0.9m/B/F3KB         \\
 1899.5923 \dotfill &  0.205 &
\phn\phs     $  60.2$ &\phn     $  -5.4$ &
     $-0.33$ &
 KPNO/0.9m/B/F3KB         \\
 1900.5674 \dotfill &  0.379 &
\phn\phs     $  46.6$ &\phn\phs $   1.5$ &
     $-0.38$ &
 KPNO/0.9m/B/F3KB         \\
 1900.5885 \dotfill &  0.383 &
\phn\phs     $  44.8$ &\phn\phs $   1.1$ &
     $-0.34$ &
 KPNO/0.9m/B/F3KB         \\
 1901.5718 \dotfill &  0.558 &
\phn         $ -41.3$ &\phn     $  -7.2$ &
     $-0.27$ &
 KPNO/0.9m/B/F3KB         \\
 2191.6387 \dotfill &  0.358 &
\phn\phs     $  56.8$ &\phn\phs $   4.9$ &
     $-0.43$ &
 DDO/1.9m/831/Thomson1KB  \\
 2192.6443 \dotfill &  0.537 &
\phn         $ -15.8$ &\phn\phs $   8.7$ &
     $-0.41$ &
 DDO/1.9m/1800/Thomson1KB \\
 2193.5913 \dotfill &  0.706 &
\nodata               &\nodata               &
     $-0.34$ &
 DDO/1.9m/1800/Thomson1KB \\
 2198.4833 \dotfill &  0.580 &
\nodata               &\nodata               &
     $-0.27$ &
 DDO/1.9m/1800/Thomson1KB \\
 2200.4987 \dotfill &  0.940 &
\nodata               &\nodata               &
     $-0.26$ &
 DDO/1.9m/1800/Thomson1KB \\
 2201.4713 \dotfill &  0.114 &
\nodata               &\nodata               &
     $-0.27$ &
 DDO/1.9m/1800/Thomson1KB \\
 2202.5888 \dotfill &  0.313 &
\nodata               &\nodata               &
     $-0.20$ &
 DDO/1.9m/1800/Thomson1KB \\
 2203.4851 \dotfill &  0.473 &
\nodata               &\nodata               &
     $-0.20$ &
 DDO/1.9m/1800/Thomson1KB \\
 2204.7099 \dotfill &  0.692 &
\phn         $ -74.8$ &\phn\phs $   2.8$ &
\phs $ 0.01$ &
 DDO/1.9m/1800/Thomson1KB \\
 2205.4764 \dotfill &  0.829 &
\phn         $ -66.9$ &\phn\phs $   6.5$ &
     $-0.21$ &
 DDO/1.9m/1800/Thomson1KB \\
 2205.5825 \dotfill &  0.848 &
\phn         $ -66.6$ &\phn\phs $   2.1$ &
\phs $ 0.02$ &
 DDO/1.9m/1800/Thomson1KB \\
 2220.4714 \dotfill &  0.507 &
\phn\phn     $  -2.5$ &\phn\phs $   7.6$ &
     $-0.44$ &
 DDO/1.9m/1800/Thomson1KB \\
 2344.9493 \dotfill &  0.735 &
\phn         $ -86.1$ &\phn     $  -3.9$ &
\phs $ 0.03$ &
 DDO/1.9m/1800/Thomson1KB \\
 2361.9052 \dotfill &  0.763 &
\phn         $ -75.4$ &\phn\phs $   6.9$ &
     $-0.47$ &
 DDO/1.9m/1800/Thomson1KB \\
 2371.9104 \dotfill &  0.550 &
\phn         $ -31.2$ &\phn     $  -0.8$ &
     $-0.17$ &
 DDO/1.9m/1800/Thomson1KB \\
 2371.9291 \dotfill &  0.553 &
\nodata               &\nodata               &
\phs $ 0.17$ &
 DDO/1.9m/1800/Thomson1KB \\
 2376.9030 \dotfill &  0.442 &
\phn\phs     $  10.0$ &         $ -10.1$ &
\phs $ 0.14$ &
 DDO/1.9m/1800/Thomson1KB \\
 2382.8795 \dotfill &  0.509 &
\phn\phn     $  -6.5$ &\phn\phs $   4.7$ &
     $-0.11$ &
 DDO/1.9m/1800/Thomson1KB \\
 2388.8805 \dotfill &  0.581 &
\phn         $ -41.7$ &\phn\phs $   1.9$ &
     $-0.06$ &
 DDO/1.9m/1800/Thomson1KB \\
 2404.6914 \dotfill &  0.404 &
\phn\phs     $  36.6$ &\phn\phs $   0.7$ &
\phs $ 0.08$ &
 DDO/1.9m/1800/Thomson1KB \\
 2404.8684 \dotfill &  0.436 &
\phn\phs     $  13.9$ &\phn     $  -8.9$ &
     $-0.13$ &
 DDO/1.9m/1800/Thomson1KB \\
 2410.7116 \dotfill &  0.479 &
\phn\phn     $  -2.1$ &\phn     $  -5.0$ &
     $-0.03$ &
 DDO/1.9m/1800/Thomson1KB \\
 2410.7443 \dotfill &  0.485 &
\phn\phn\phs $   4.5$ &\phn\phs $   4.3$ &
     $-0.21$ &
 DDO/1.9m/1800/Thomson1KB \\
 2413.7035 \dotfill &  0.013 &
\phn\phn\phs $   6.1$ &\phn\phs $   6.7$ &
     $-0.18$ &
 DDO/1.9m/1800/Thomson1KB \\
 2414.7121 \dotfill &  0.194 &
\phn\phs     $  65.4$ &\phn\phs $   1.5$ &
\phs $ 0.02$ &
 DDO/1.9m/1800/Thomson1KB \\
 2414.7522 \dotfill &  0.201 &
\phn\phs     $  65.8$ &\phn\phs $   0.8$ &
     $-0.01$ &
 DDO/1.9m/1800/Thomson1KB \\
 2414.7836 \dotfill &  0.206 &
\phn\phs     $  69.4$ &\phn\phs $   3.6$ &
     $-0.10$ &
 DDO/1.9m/1800/Thomson1KB \\
 2414.8197 \dotfill &  0.213 &
\phn\phs     $  70.7$ &\phn\phs $   4.2$ &
     $-0.23$ &
 DDO/1.9m/1800/Thomson1KB \\
\enddata
\end{deluxetable}

\clearpage

% Table 2 - Orbital Elements

\begin{deluxetable}{lcc}
\tablewidth{0pc}
\tablenum{2}
\tablecaption{Circular Orbital Elements\label{tab2}}
\tablehead{
\colhead{Element}         &\colhead{Brocksopp et al. (1999)} &\colhead{This Work}   }
\startdata
$P$ (d)                   & 5.599829 (16)      &  5.599829\tablenotemark{a} \\
$T$(IC) (HJD)             & 2,441,874.707 (9)  &  2,451,730.449 (8)  \\
$K_1$ (km s$^{-1}$)       & 74.9 (6)           &  75.6 (7)           \\
$V_0$ (km s$^{-1}$)       & \nodata            &  $-$7.0 (5)         \\
$\sigma$ (km s$^{-1}$)    & \nodata            &  5.3                \\
$f(m)$ ($M_\odot$)        & 0.244              &  0.251 (7)          \\
$a_1 \sin i$ ($R_\odot$)  & \nodata            &  8.36 (8)           \\
\enddata
\tablenotetext{a}{Fixed.}
\tablecomments{Numbers in parentheses give the error in the last digit quoted.}
\end{deluxetable}

\newpage

%%%%%%%%%%%%%%%%%%%%%%%%%%%%%%%%%%%%%%%%%%%%%%%%%%%%%%%%%%%%%%%

% Figures

\clearpage

\setcounter{figure}{0}

% Figure 1
\begin{figure}
\plotone{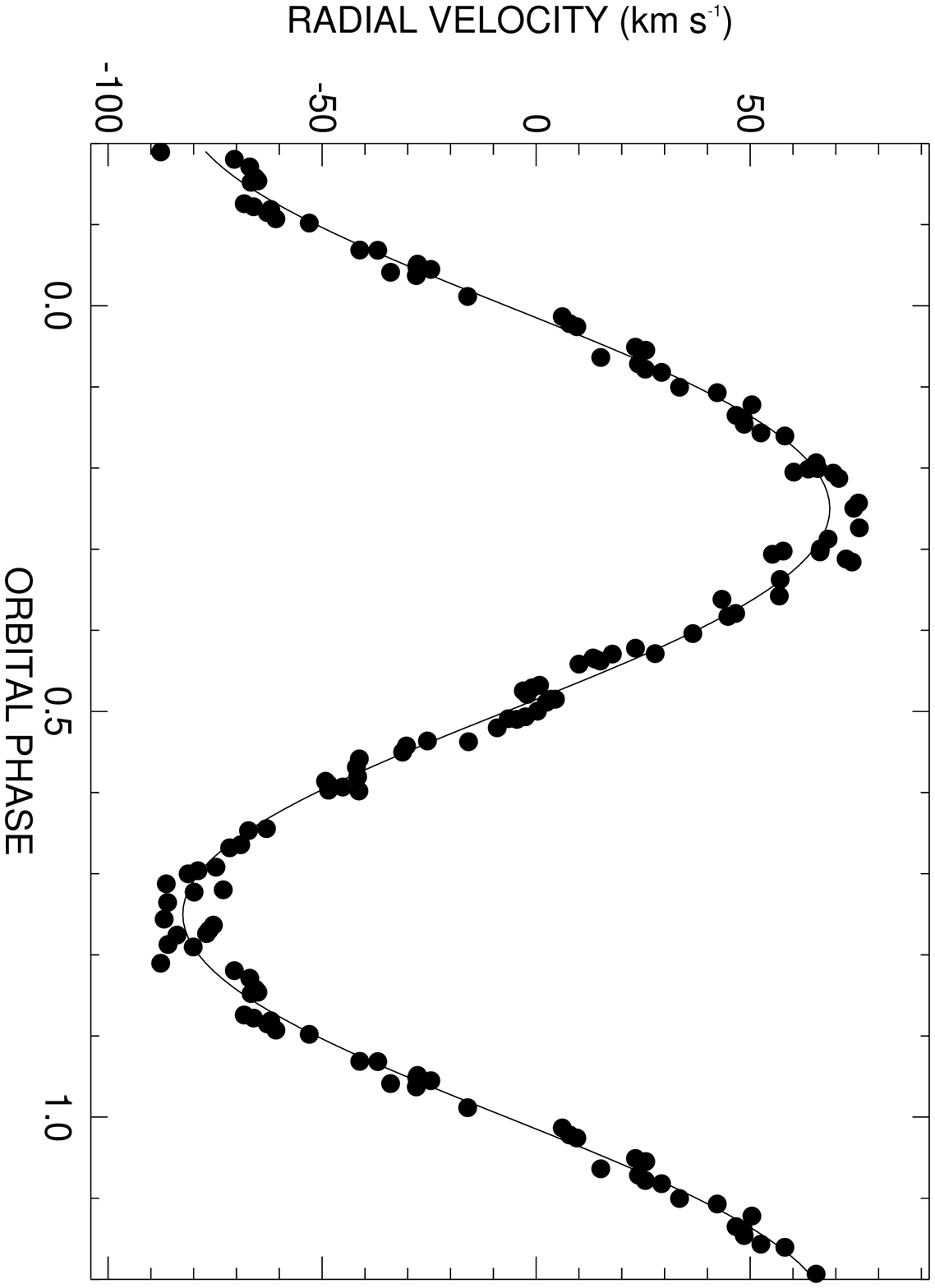}
\caption{}
\end{figure}

% Figure 2
\begin{figure}
\plotone{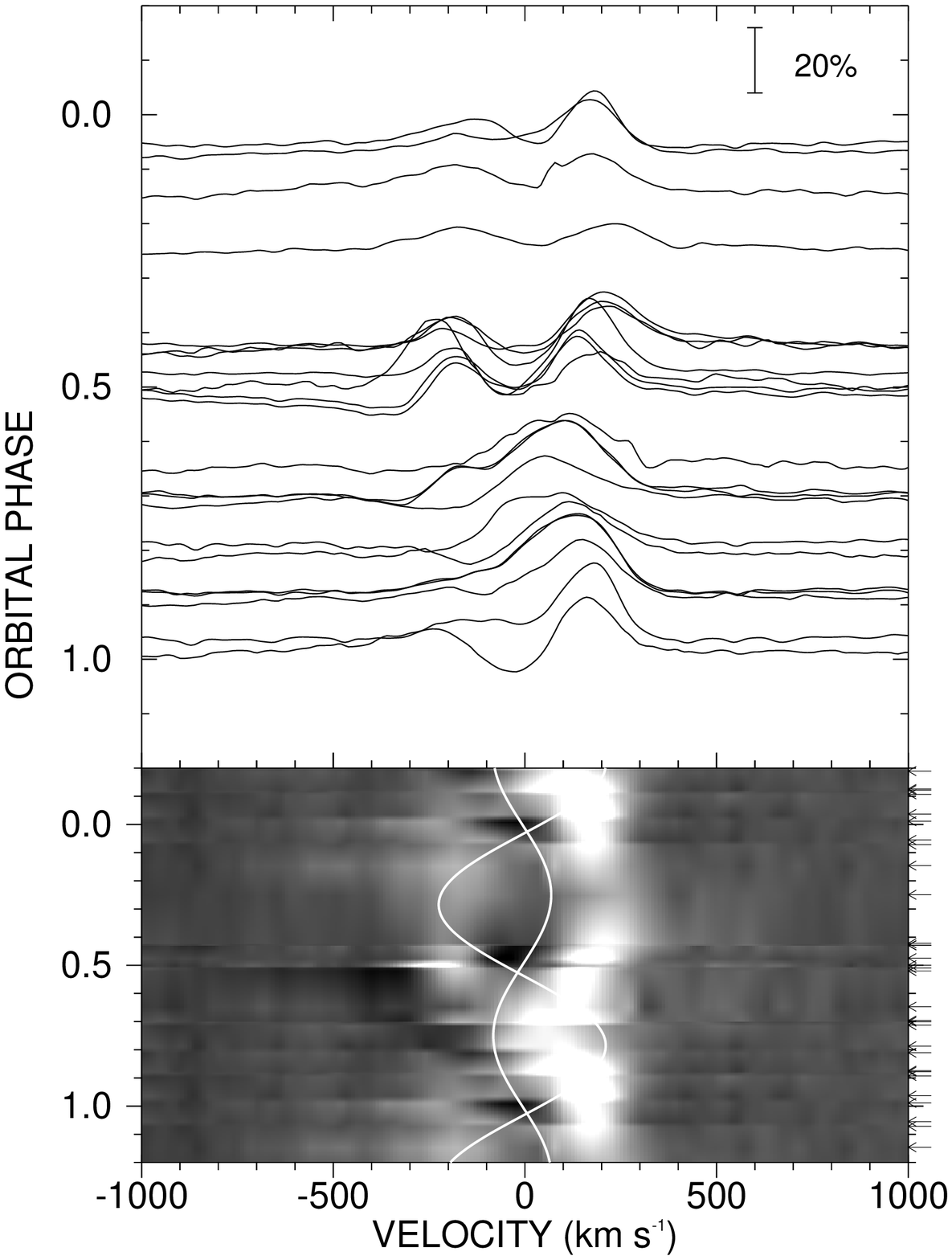}
\caption{}
\end{figure}

% Figure 3
\begin{figure}
\plotone{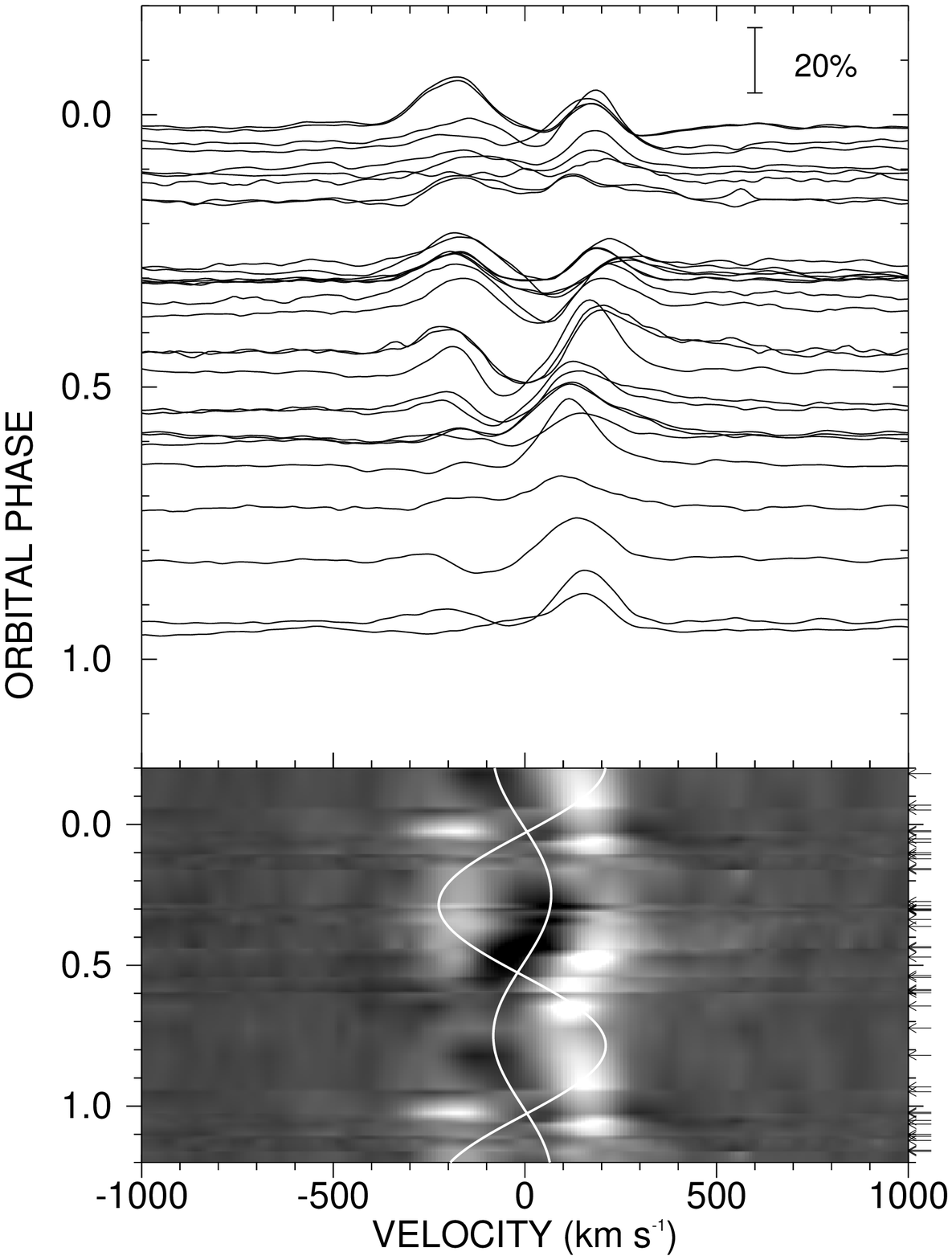}
\caption{}
\end{figure}

% Figure 4
\begin{figure}
\plotone{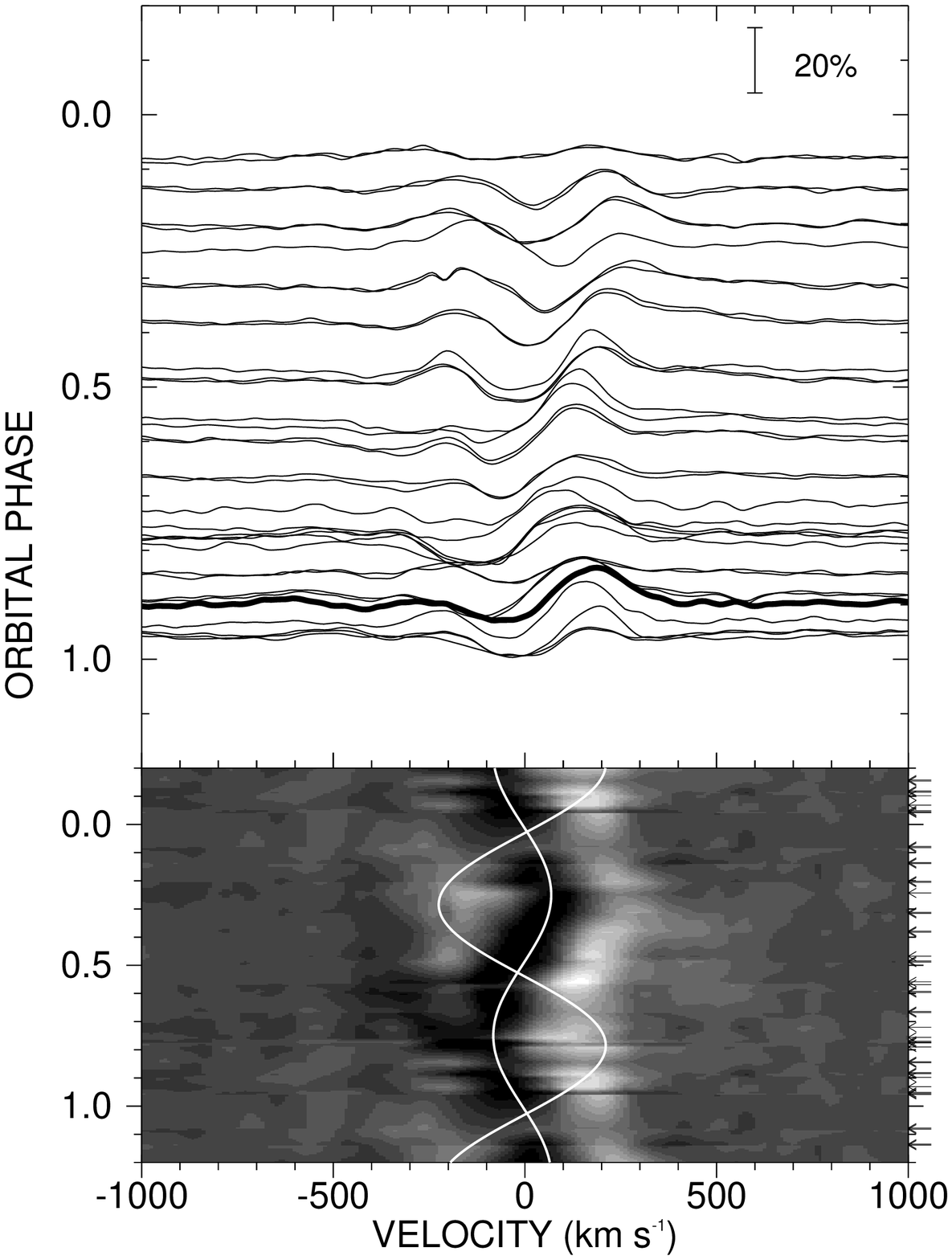}
\caption{}
\end{figure}

% Figure 5
\begin{figure}
\plotone{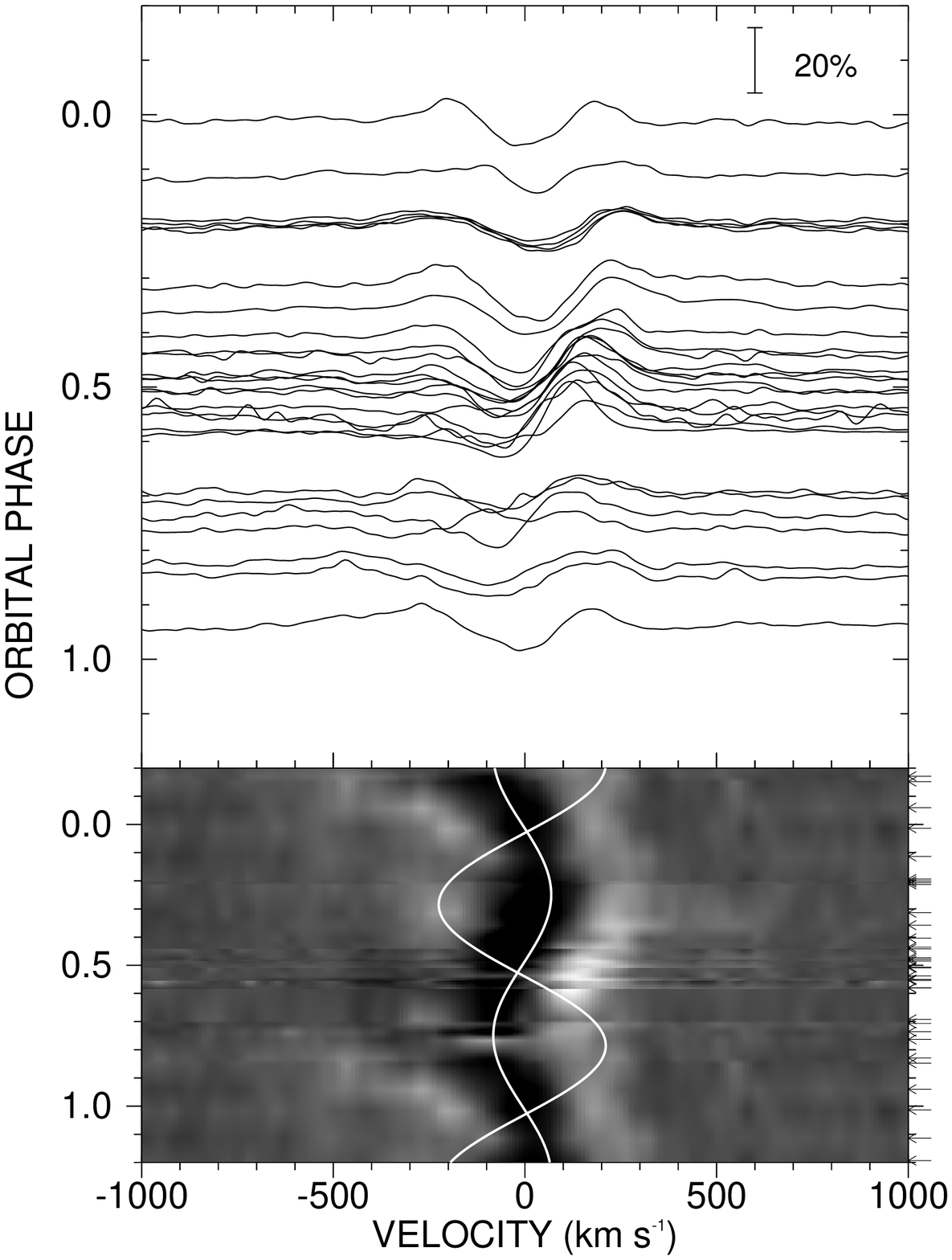}
\caption{}
\end{figure}

% Figure 6
\begin{figure}
\plotone{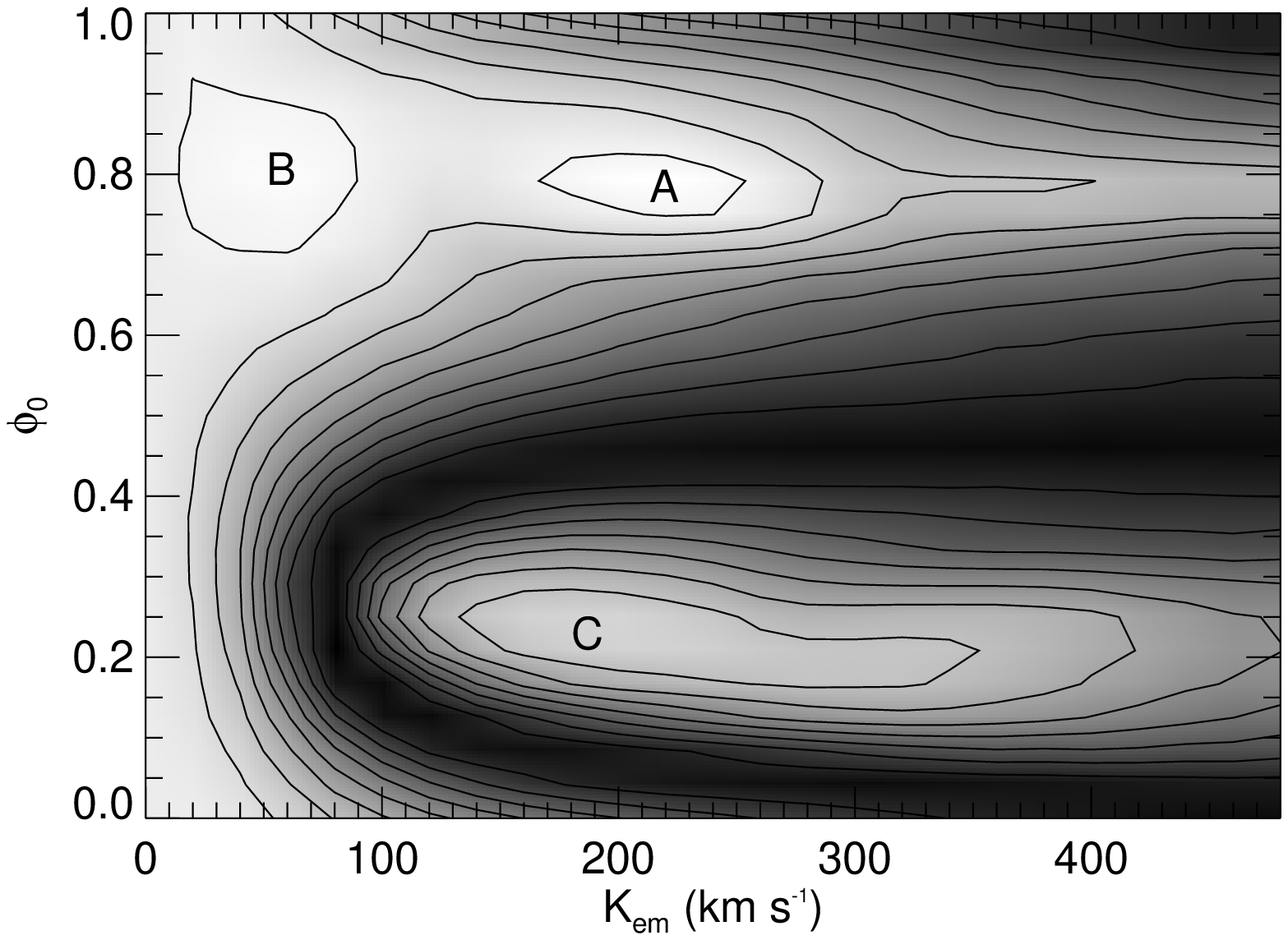}
\caption{}
\end{figure}

% Figure 7
\begin{figure}
\plotone{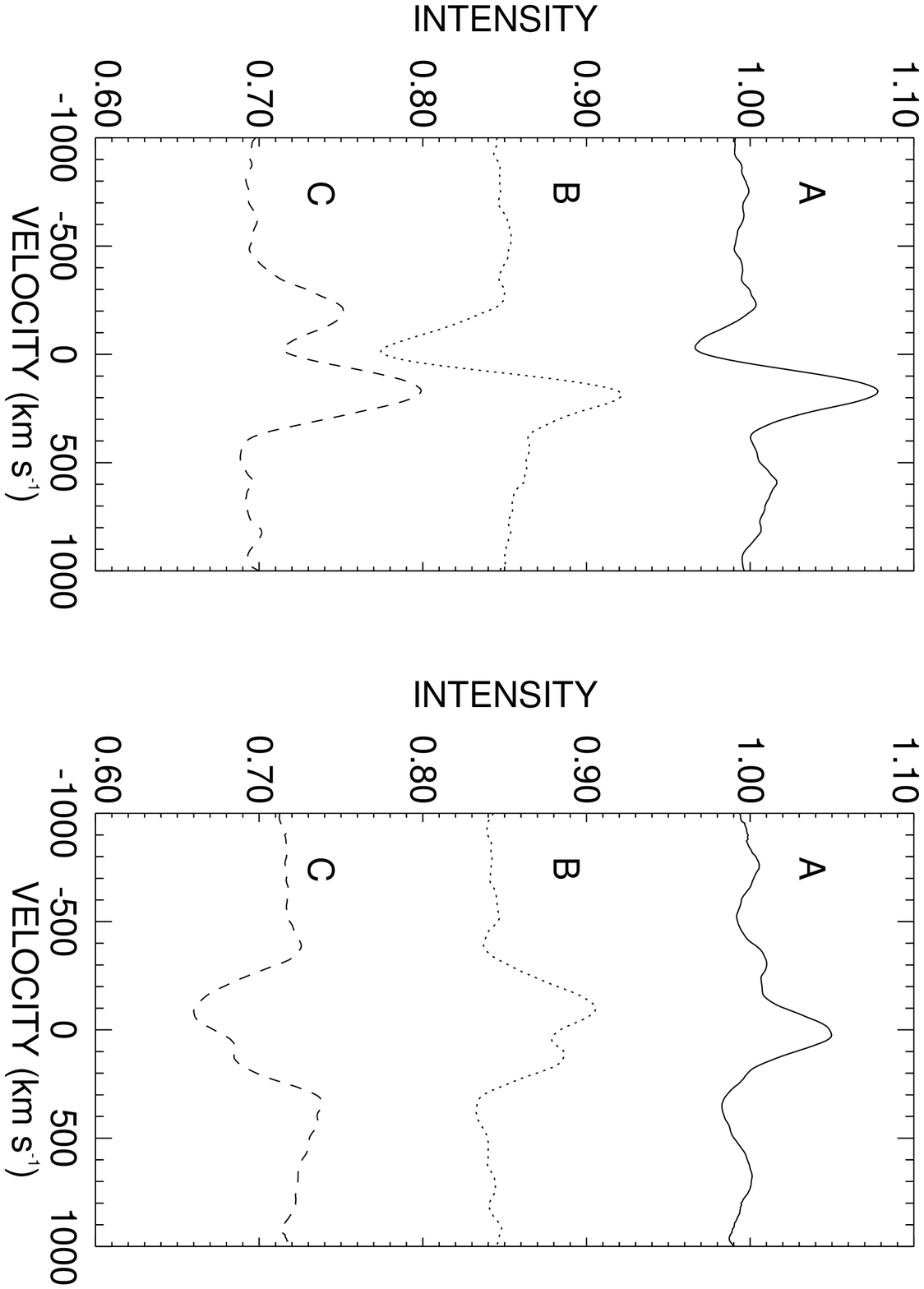}
\caption{}
\end{figure}

% Figure 8
\begin{figure}
\plotone{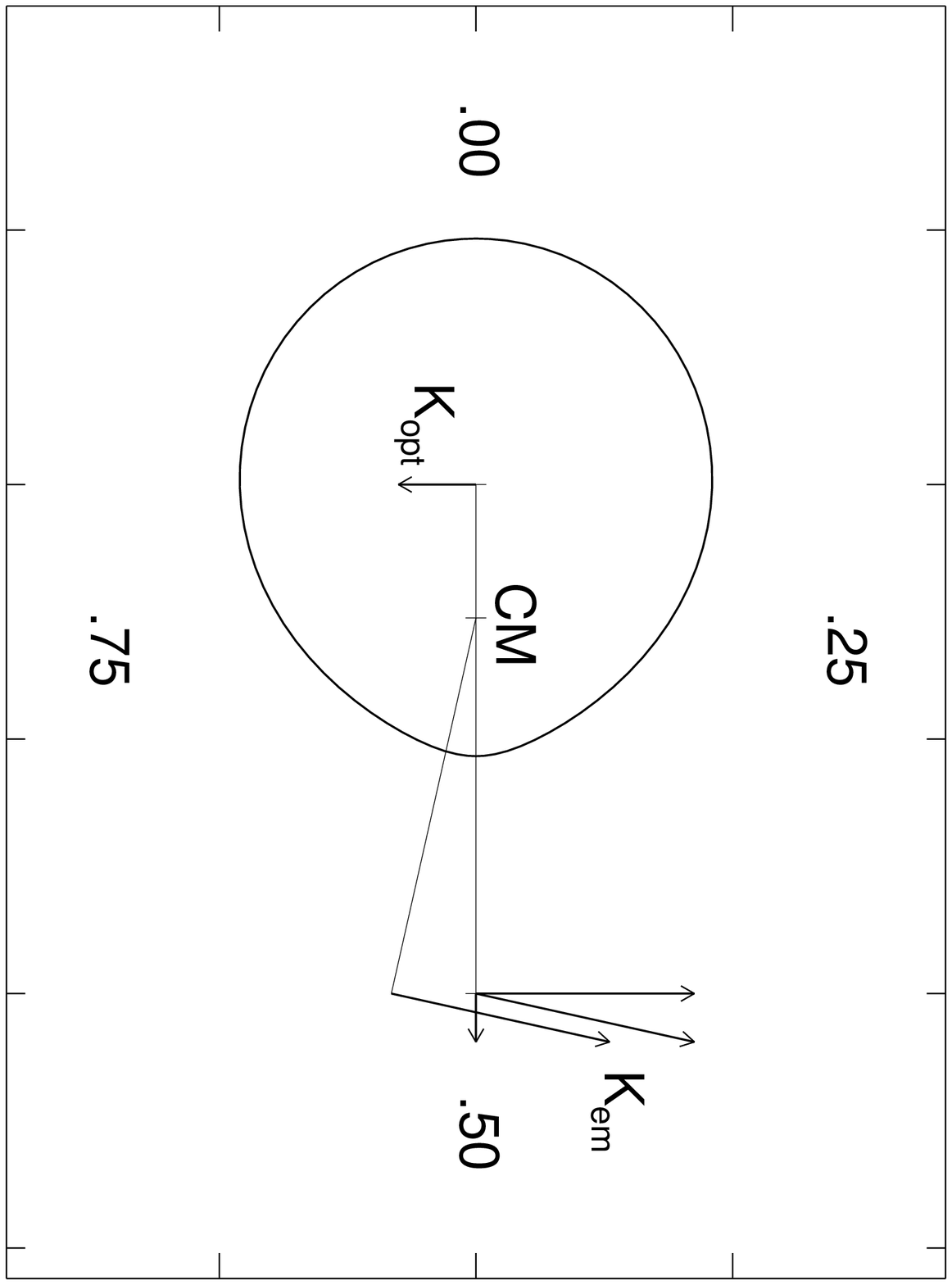}
\caption{}
\end{figure}

% Figure 9
\begin{figure}
\plotone{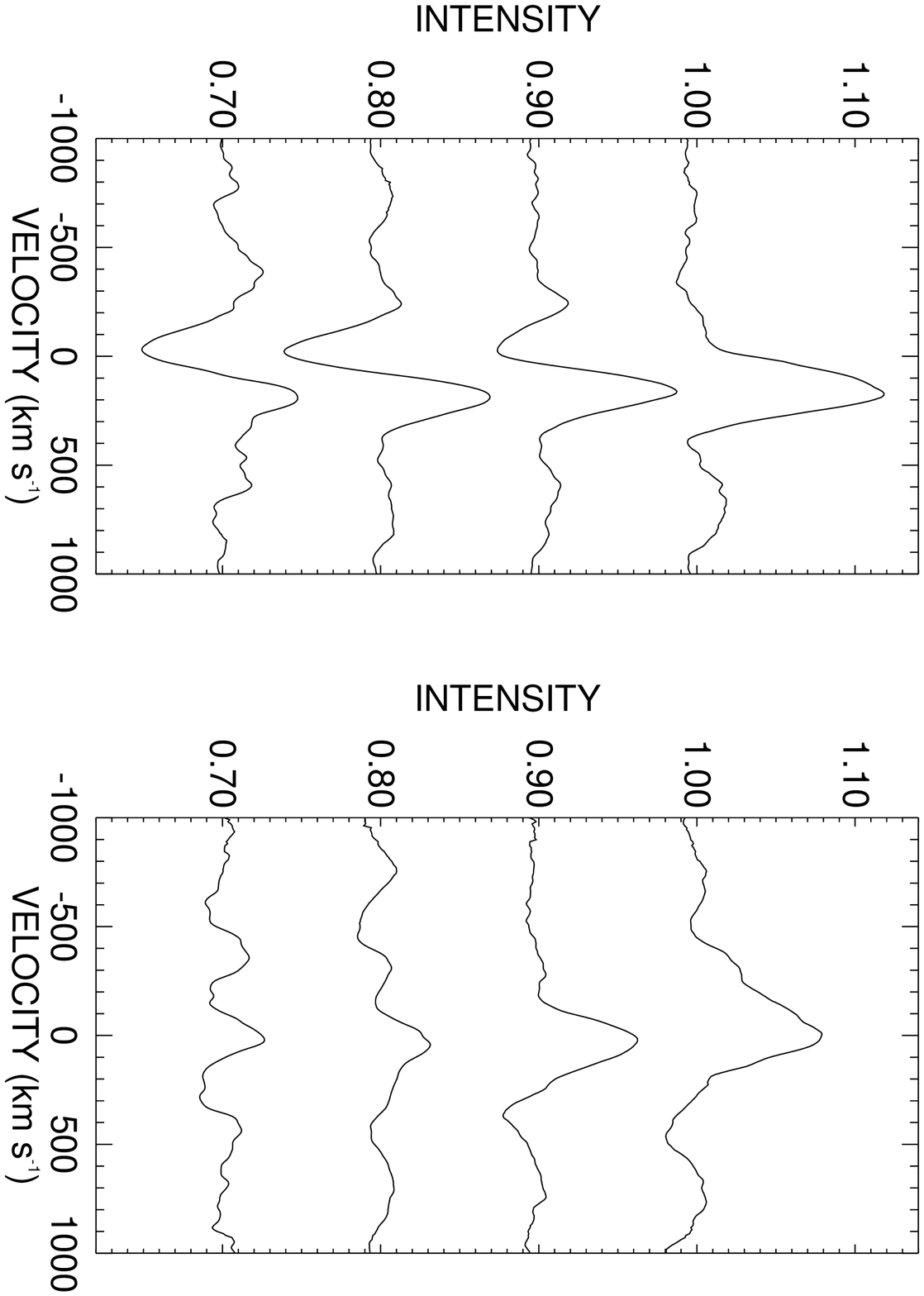}
\caption{}
\end{figure}

% Figure 10
\begin{figure}
\plotone{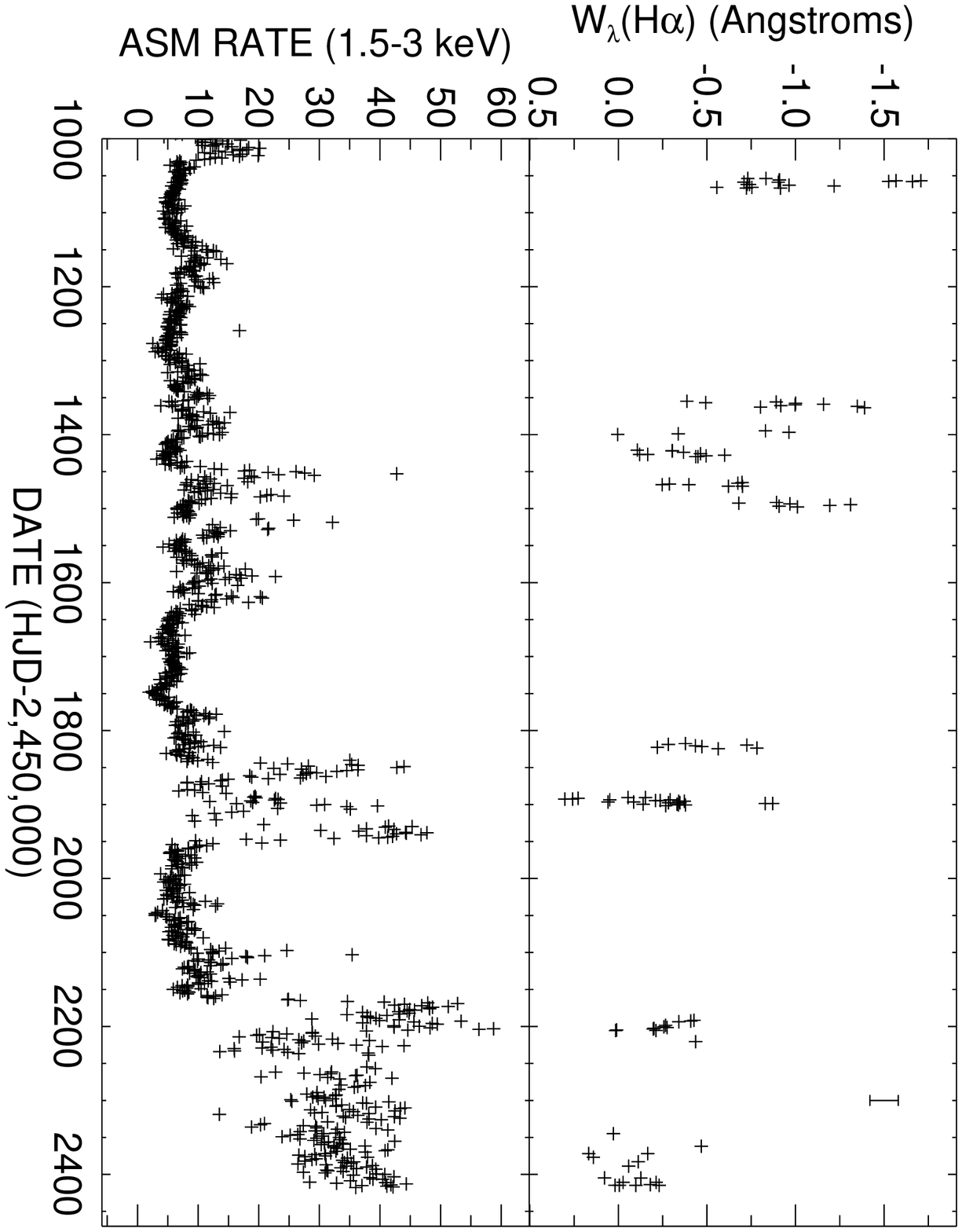}
\caption{}
\end{figure}

% Figure 11
\begin{figure}
\plotone{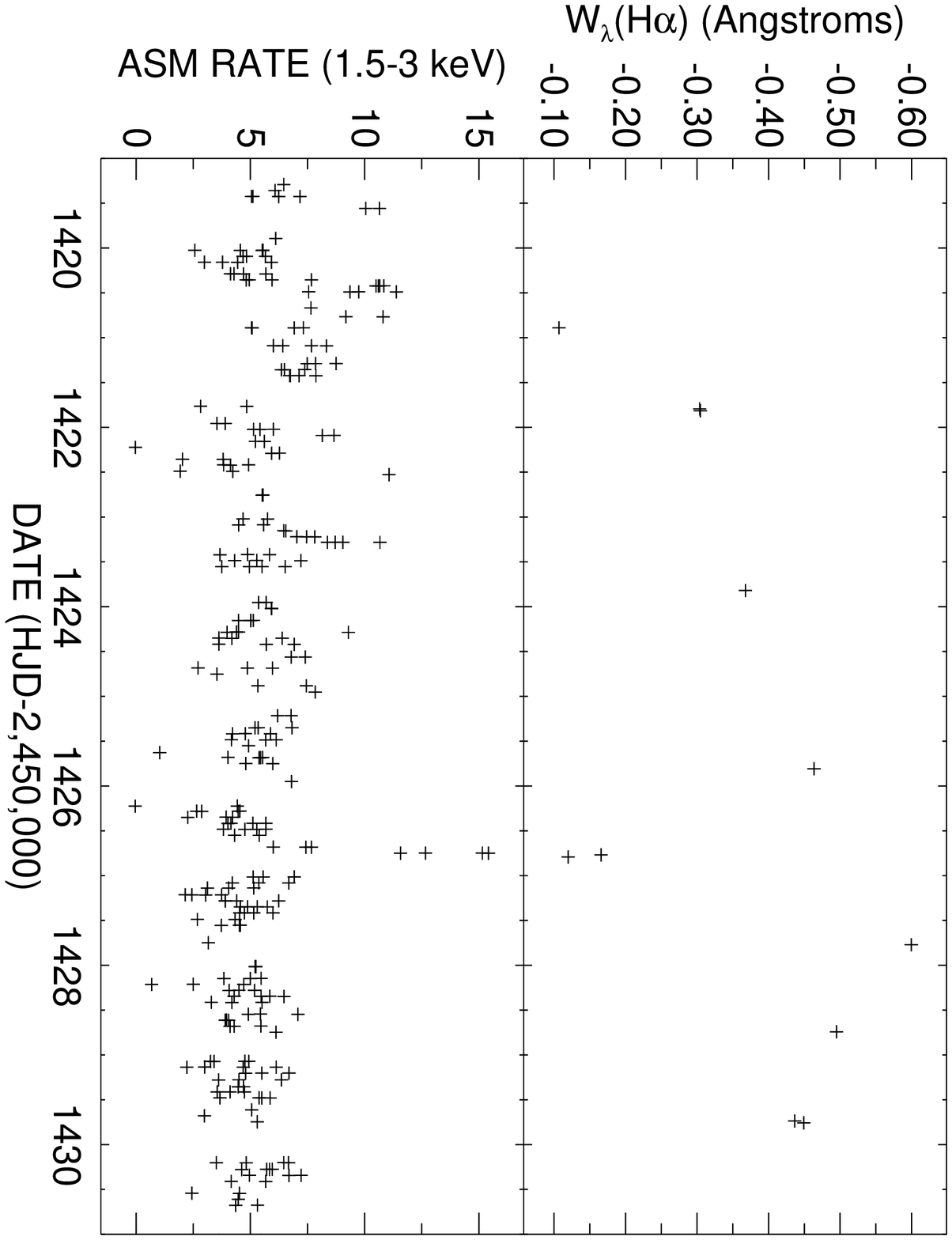}
\caption{}
\end{figure}

% Figure 12
\begin{figure}
\plotone{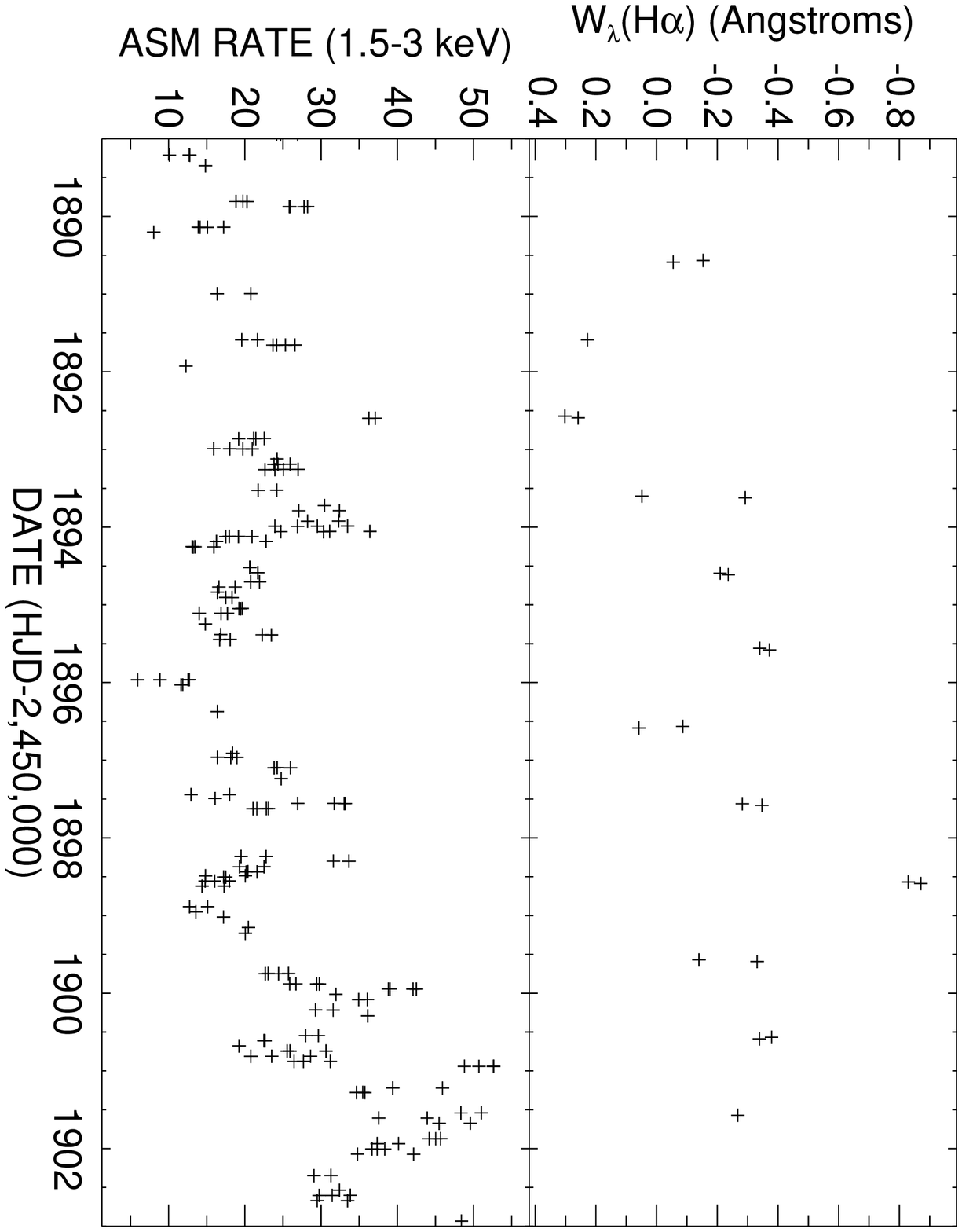}
\caption{}
\end{figure}

% Figure 13
\begin{figure}
\plotone{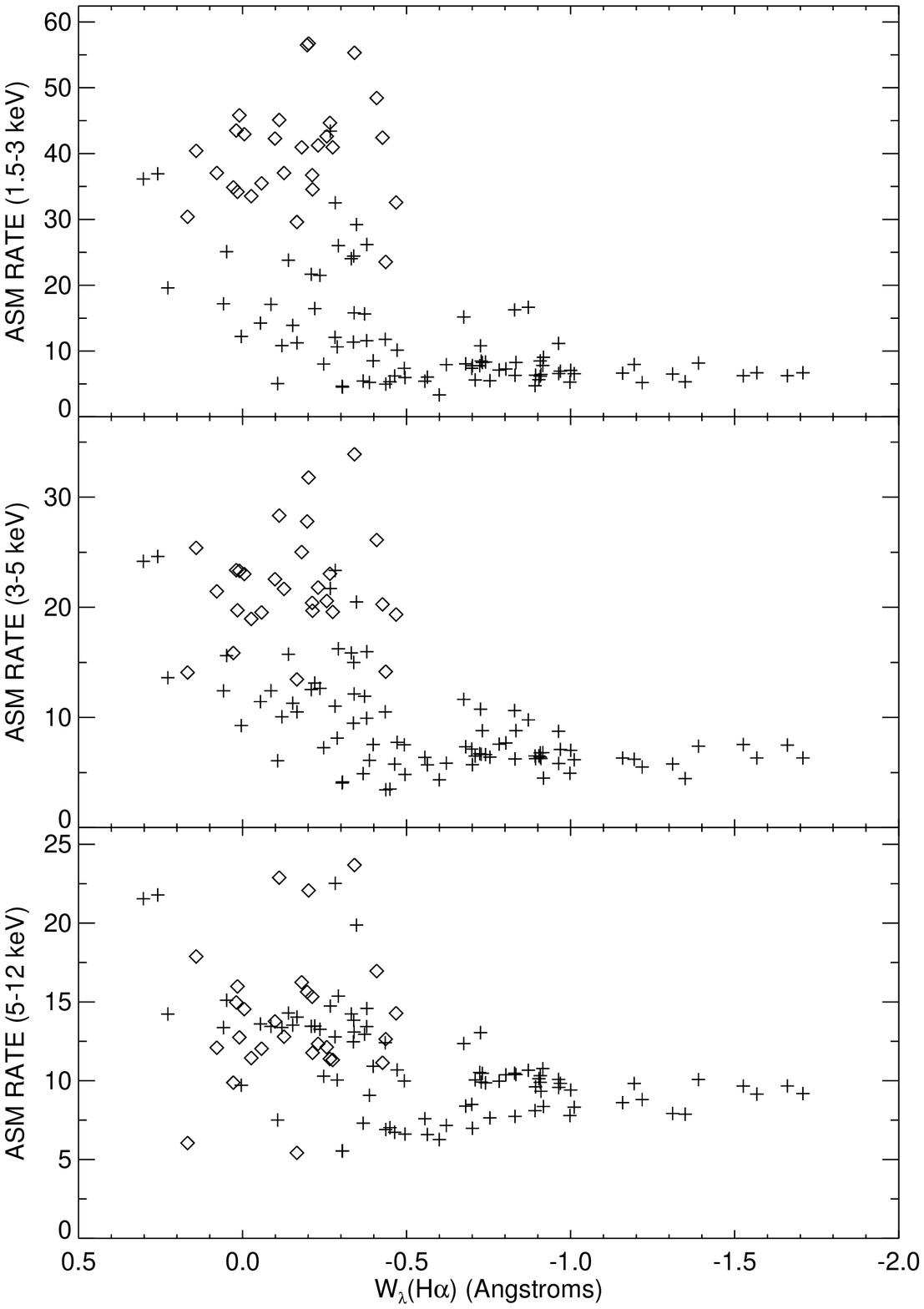}
\caption{}
\end{figure}

% Figure 14
\begin{figure}
\plotone{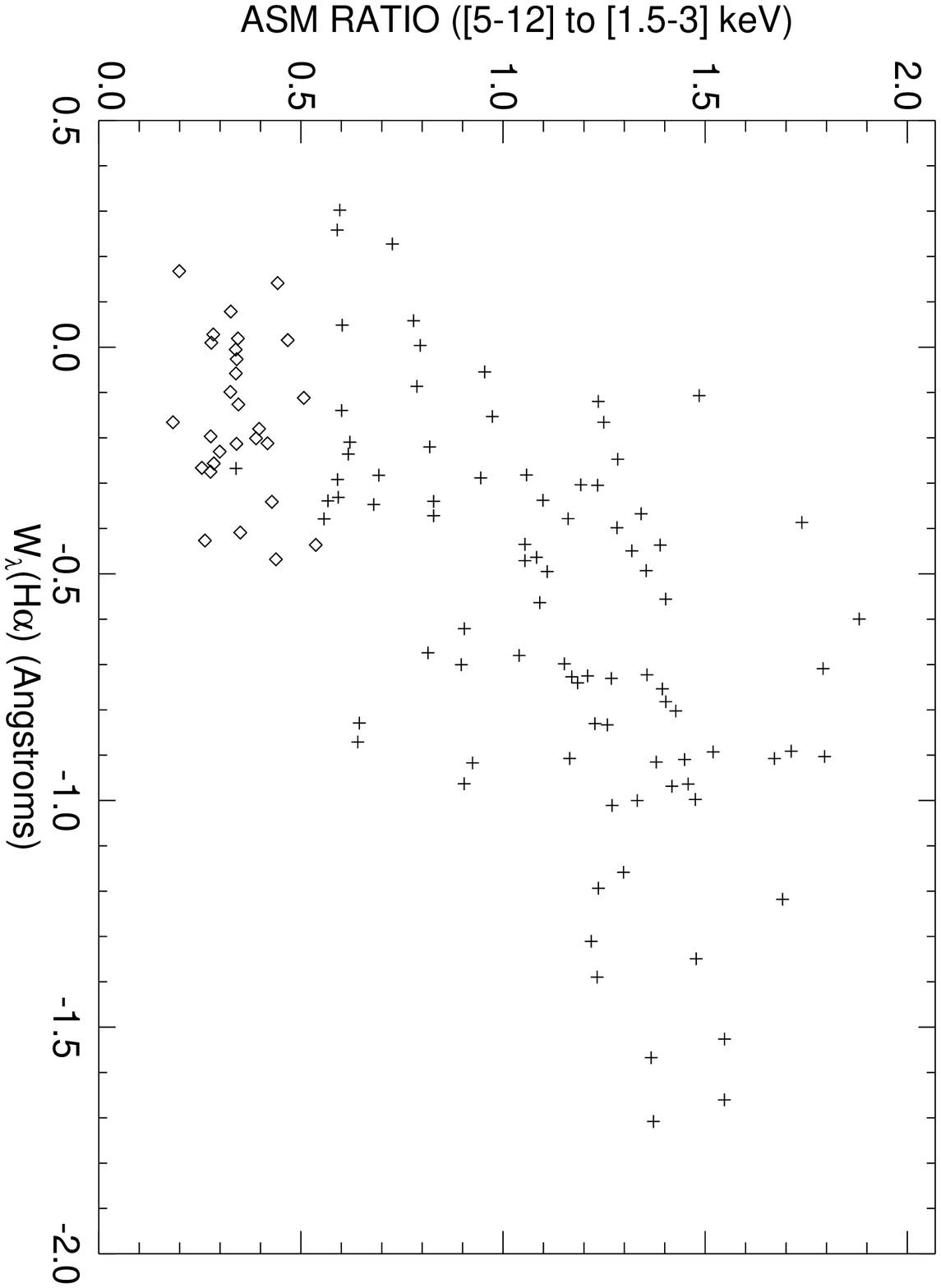}
\caption{}
\end{figure}

% Figure 15
\begin{figure}
\plotone{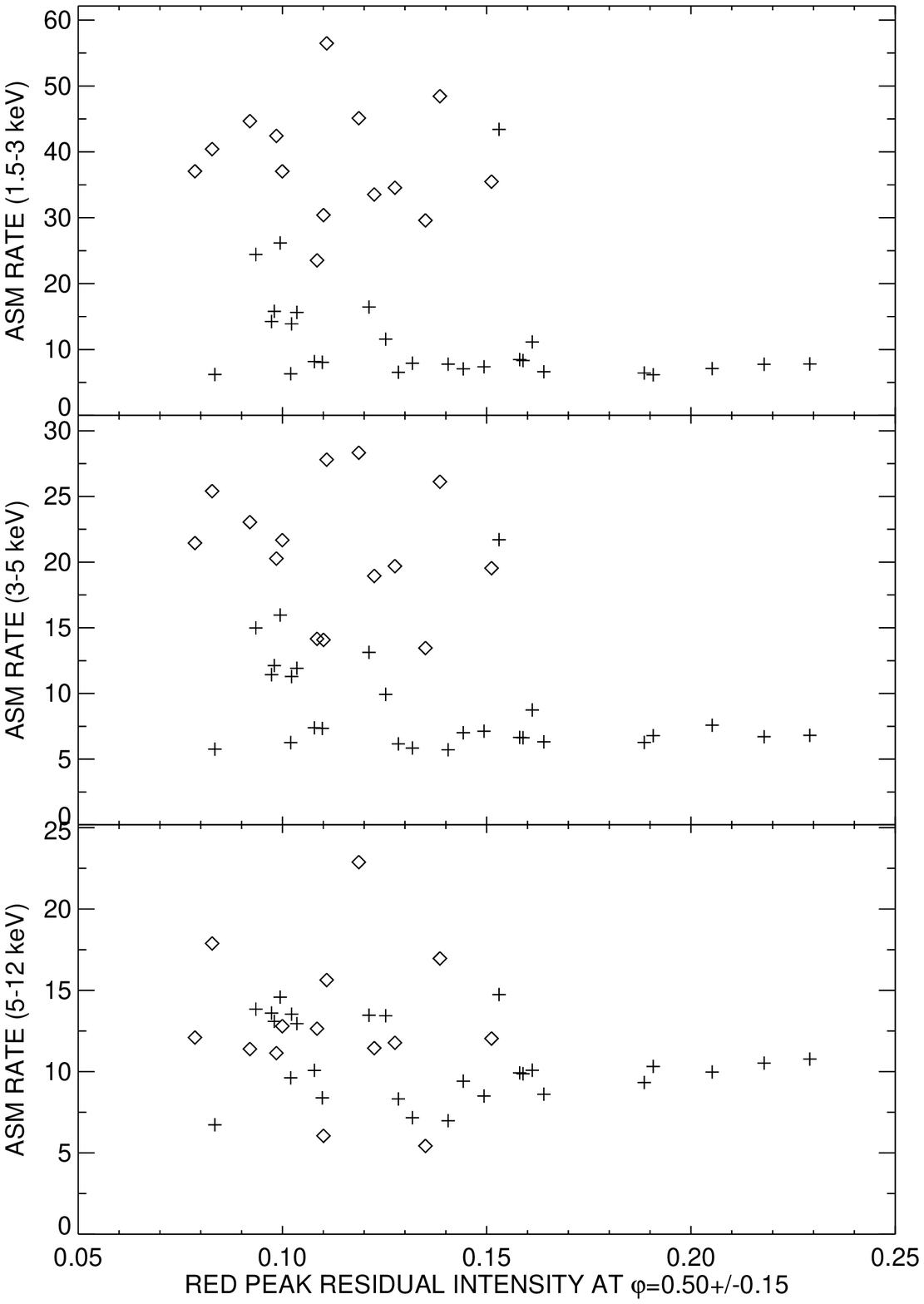}
\caption{}
\end{figure}

% Figure 16
\begin{figure}
\plotone{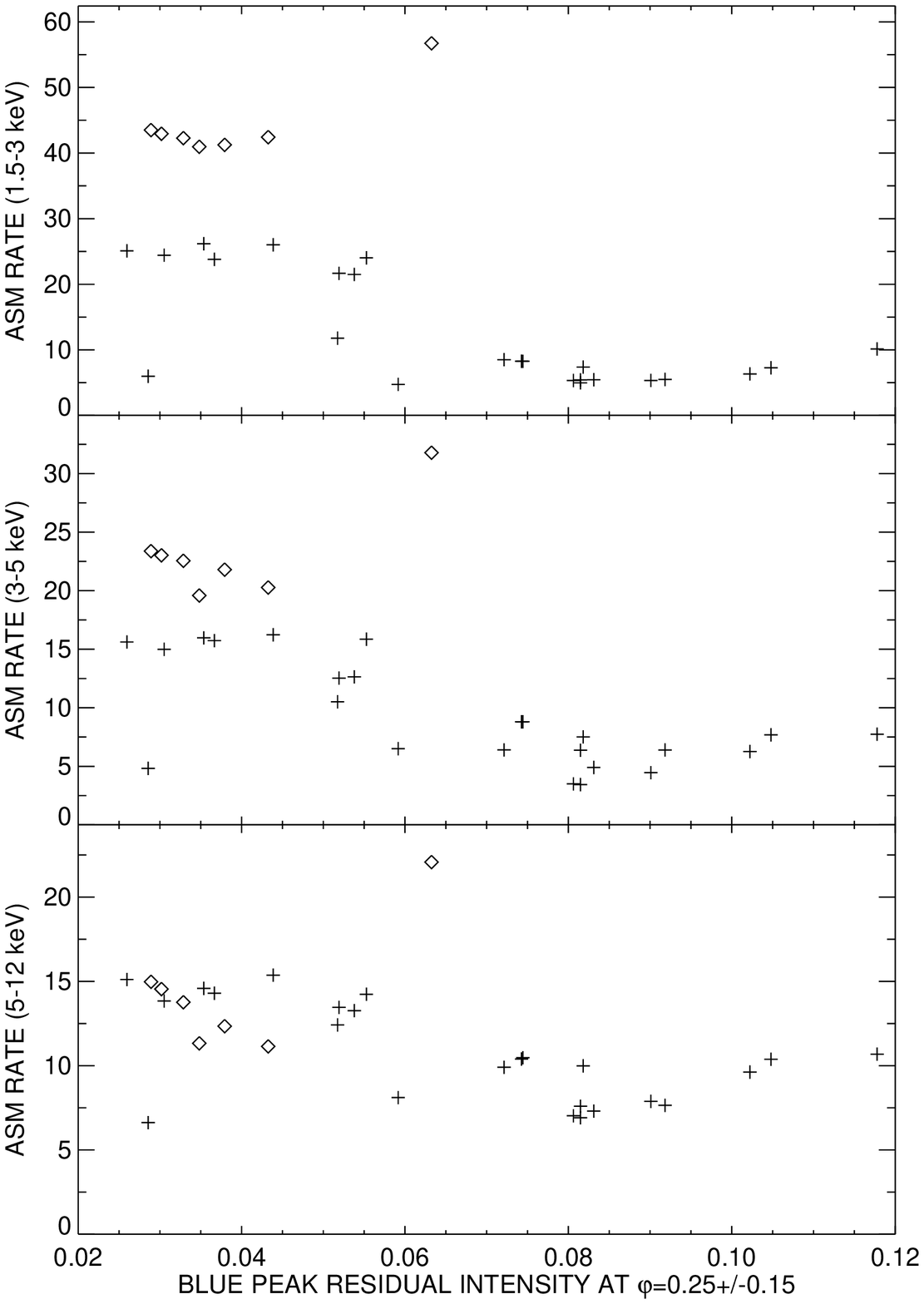}
\caption{}
\end{figure}

%%%%%%%%%%%%%%%%%%%%%%%%%%%%%%%%%%%%%%%%%%%%%%%%%%%%%%%%%%%%%%%

\end{document}